\documentclass[aps,prb,twocolumn,superscriptaddress]{revtex4-1}
\usepackage{amsmath}
\usepackage{amssymb}
\usepackage{graphicx}
\usepackage{color}

\usepackage{hyperref}
\usepackage[caption = false]{subfig}
\usepackage[export]{adjustbox}

\begin{document}

\title{Spin-density-wave order controlled by uniaxial stress in CeAuSb$_2$}

\author{R. Waite}
\affiliation{H. H. Wills Physics Laboratory, University of Bristol, Bristol, BS8 1TL, United Kingdom.}
\affiliation{ ISIS Facility, Rutherford Appleton Laboratory, Chilton, Didcot, OX11 0QX, United Kingdom.}
\author{F. Orlandi}
\affiliation{ ISIS Facility, Rutherford Appleton Laboratory, Chilton, Didcot, OX11 0QX, United Kingdom.}
\author{D. A. Sokolov}
\affiliation{Max Planck Institute for Chemical Physics of Solids, N{\"o}thnitzer Stra{\ss}e 40, 01187 Dresden, Germany.}
\author{R.A. Ribeiro}
\affiliation{Ames Laboratory, U.S. DOE, and Department of Physics and Astronomy,Iowa State University, Ames, Iowa 50011, United States.}
\author{P. C. Canfield}
\affiliation{Ames Laboratory, U.S. DOE, and Department of Physics and Astronomy,Iowa State University, Ames, Iowa 50011, United States.}
\author{P. Manuel}
\affiliation{ ISIS Facility, Rutherford Appleton Laboratory, Chilton, Didcot, OX11 0QX, United Kingdom.}
\author{D. D. Khalyavin}
\affiliation{ ISIS Facility, Rutherford Appleton Laboratory, Chilton, Didcot, OX11 0QX, United Kingdom.}
\author{C. W. Hicks}
\affiliation{Max Planck Institute for Chemical Physics of Solids, N{\"o}thnitzer Stra{\ss}e 40, 01187 Dresden, Germany.}
\affiliation{School of Physics and Astronomy, University of Birmingham, Birmingham B15 2TT}
\author{S. M. Hayden}
\affiliation{H. H. Wills Physics Laboratory, University of Bristol, Bristol, BS8 1TL, United Kingdom.}

\begin{abstract}  
The tetragonal heavy-fermion compound CeAuSb$_2$ (space group $P4/nmm$) exhibits incommensurate spin density wave (SDW) order below $T_{N}\approx6.5~K$ with the propagation vector $\mathbf{q}_A = (\delta_A,\delta_A,1/2)$. 
The application of uniaxial stress along the [010] direction induces a sudden change in the resistivity ratio $\rho_a/\rho_b$ at a compressive strain of $\epsilon \approx -0.5$\%.
Here we use neutron scattering to show that the uniaxial stress induces a first-order transition to a SDW state with a different propagation vector $(0,\delta_B,1/2)$ with $\delta_B=0.25$.  The magnetic structure of the new (B) phase consists of Ce layers with ordered moments alternating with layers with zero moment stacked along the $c$-axis. The ordered layers have an up-up-down-down configuration along the $b$-axis. This is an unusual situation in which the loss of spatial inversion is driven by the magnetic order. We argue that the change in SDW wavevector leads to  Fermi surface reconstruction and a concomitant change in the transport properties.

\end{abstract}

\maketitle
\setlength{\parskip}{8pt}

\section{Introduction}

Heavy fermions systems \cite{Fisk1986,Knebel2001} are metals incorporating elements with partially filled 4$f$ or 5$f$ shells with heavy electron quasiparticles which show masses of up to 100 times those in a conventional metal. Our understanding of cerium-based heavy-fermion materials is based on the Kondo lattice model\cite{Doniach1977}.  Ce 4$f$ electron spins are localised at high temperatures leading to a small Fermi surface. As the temperature is lowered, the $f$ electrons are screened by other electrons, below a characteristic Kondo temperature $T_K$ they become itinerant and the Fermi surface volume increases. The magnetism in heavy fermion systems is described by a Ruderman–Kittel–Kasuya–Yosida (RKKY) exchange interactions between localised moments or by a wavevector dependent susceptibility derived from the electronic band structure involving the itinerant $f$ electron states. 

Heavy fermions (and other strongly correlated systems) are delicately balanced systems with competing interactions which can be perturbed by hydrostatic pressure, uniaxial stress or magnetic field. Hydrostatic pressure increases the overlap and hybridisation of the atomic orbitals and can lead to a transition from the localised to delocalised $f$ electrons \cite{Shishido2005} and a corresponding collapse of magnetic order \cite{Knebel2001}.  Heavy fermion systems can also be very sensitive to the application of a magnetic field. For example, for CeAuSb$_2$ \cite{Balicas2005, Zhao2016, Marcus2018}, applying a field causes a transition between magnetically ordered states and the suppression of magnetic order. More recently, resistivity measurements under uniaxial stress \cite{Park2018,Park2018a} have revealed an additional phase transition in CeAuSb$_2$.  Here we use neutron scattering to show that this is a transition to a new magnetic state and determine the nature of the order.   

\begin{figure}[!h]
\centering
\includegraphics[width=0.8\linewidth]{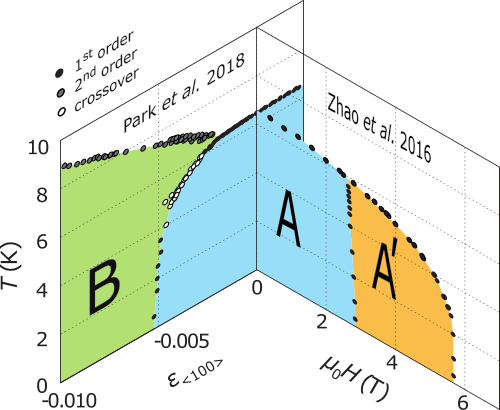}
\caption{Phase diagram of CeAuSb$_2$ determined from transport measurements in Refs.~\onlinecite{Zhao2016,Park2018,Park2018a} for a magnetic field applied along the $[001]$ and a compressive strain $\epsilon_{\langle 100 \rangle}$ along the $\langle 100 \rangle$-type direction.} 
\label{fig:CeAuSb2_PhaseDiag}
\end{figure}

CeAuSb$_2$ is a heavy-fermion antiferromagnet \cite{Thamizhavel2003,Balicas2005} with linear coefficient of specific heat $\gamma \approx 0.5$~JK$^{-2}$~mole$^{-1}$, a Kondo temperature of $\sim$14~K and SDW  order below $T_{\textrm{SDW}}\approx 6.5$~K at zero stress and magnetic field.  
It crystallises with a tetragonal space group $P4/nmm$ in a quasi-2D structure that consists of alternating CeSb-Au and CeSb-Sb planes stacked along the $c$-axis and the in-plane Ce-Ce bonds are along $a/b$ axes.  The application of a magnetic field \cite{Balicas2005, Zhao2016} along the $c$-axis
leads to a magnetisation curve $M(B)$ with metamagnetic anomalies and resistivity anomalies at $\mu_0 B_1=2.78$~T and $\mu_0 B_2=5.42$~T. These anomalies can be used to identify two phases A and A$^{\prime}$ as shown in Fig.~\ref{fig:CeAuSb2_PhaseDiag}.  
Neutron diffraction measurements by Marcus \textit{et al.}\cite{Marcus2018} show that the A phase is a SDW state with ordering wavevectors $\mathbf{q}_{1,2} = (\delta_A,\pm\delta_A,0.5)$, where $\delta = 0.136$, and an ordered-moment polarised along the $c$-axis. The absence of third harmonics of $\mathbf{q}_{1}$ led Marcus \textit{et al.} to conclude that the magnetism in CeAuSb$_2$ requires an itinerant description of the $f$ electron states in contrast to systems such as CeSb  \cite{Chattopadhyay1994} which show strong third harmonics and require a RKKY model.  Marcus \textit{et al.} found that the A phase has single-$\mathbf{q}$ SDW domains with $\mathbf{q}_{1}$ or $\mathbf{q}_{2}$ order and microscopic orthorhombic symmetry.  In contrast, the A$^{\prime}$ is found to have a multi-$\mathbf{q}$ structure in which modulations with wavevectors $\mathbf{q}_{1}$, $\mathbf{q}_{2}$, and $\mathbf{q}_{1} \pm \mathbf{q}_{2}$ co-exist.  

The fact that a modest magnetic field can induce changes in the structure of the SDW order suggests that CeAuSb$_2$ may have multiple nearly-degenerate ordered states. Another useful perturbation or tuning parameter is uniaxial stress applied, for example, along the $\langle 100 \rangle$-type lattice direction of the tetragonal parent structure. In this paper we designate the axis along which stress is applied as the $b$-axis, or $[010]$.
In contrast to applying the field along $[001]$, this perturbation breaks the tetragonal lattice symmetry. The temperature/strain/magnetic field diagram of CeAuSb$_2$ has been extensively studied by transport and heat capacity~\cite{Zhao2016,Park2018,Park2018a}  (see Fig.~\ref{fig:CeAuSb2_PhaseDiag} for a summary).
Stress applied along $[010]$ induces a transition into a new ``B phase'' at a strain of $\epsilon \sim -0.5$\% (where negative $\epsilon$ denotes compressive strain) at the lowest temperatures. 
The transition to the B phase is characterised by a sudden jump in the resistivity which suggests a first order transition, in particular, the resistivity along the $[100]$ is enhanced. An analogous anisotropy in resistivity is induced in the SDW phases of Sr$_3$Ru$_2$O$_7$ when a component of magnetic field is applied along $[100]$ \cite{Borzi2007,Lester2015}.

In order to understand the unusual transport properties of the CeAuSb$_2$ B phase it is important to establish the nature of the SDW order. Here we report neutron diffraction measurements performed with in-situ uniaxial stress. Our measurements reveal that the B phase has a single-$\mathbf{q}$ SDW order with the moment polarised along the $c$-axis and propagation vector, $\mathbf{q}_B = (0, 0.25, 0.5)$. 
The most obvious mechanism responsible for the anisotropic resistivity is a Fermi surface reconstruction caused by the SDW. 
In addition, the phase diagram of CeAuSb$_2$ has a strong similarity to those of the other heavy-fermion systems CeNiGe$_3$~\cite{Mun2010}, CeRh$_2$Si$_2$~\cite{Knafo2010} and YbNiSi$_3$~\cite{Bud2007}, which suggests that the phenomenology we will describe here might be rather common. In all these compounds, the easy axis is the $c$-axis, and transition from the zero-field magnetic order to a homogeneously-polarized state occurs via two first-order metamagnetic transitions, through an intermediate phase with higher resistivity.

\section{Experiment}

\subsection{Sample Growth and Characterisation}

\begin{figure}[!htb]
\centering
\includegraphics[width=0.7\linewidth]{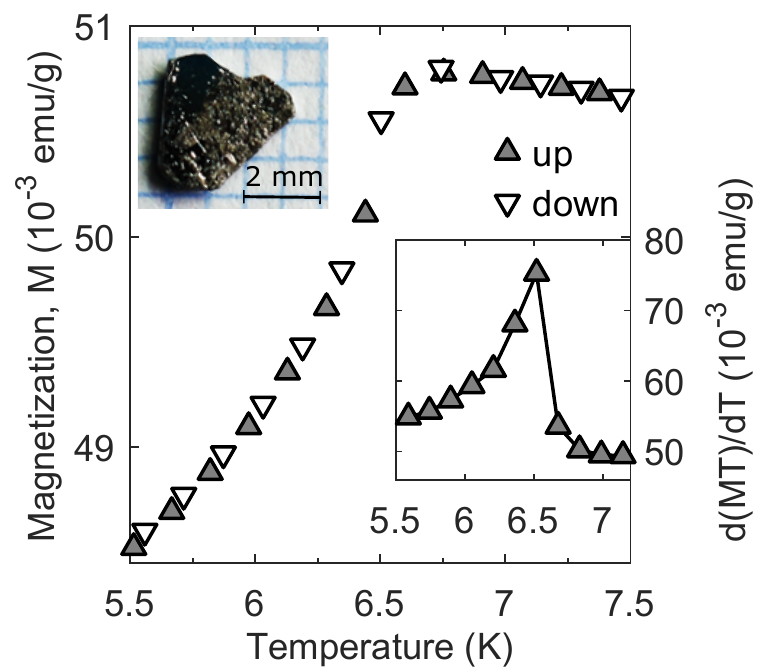}
\caption{DC SQUID magnetisation measurements for up and down temperature sweeps for the sample measured with neutron diffraction (shown in inset). A field of 0.5~T was applied in the $ab$-plane. The inset plot shows the derivative $d(MT)/dT$ for the up-sweep data - for an antiferromagnet at constant field this quantity is proportional to the heat capacity, $C_p(T)$ \cite{Fisher1962}. The transition into the SDW-A phase occurs at $T_\mathrm{N}\approx6.5$~K.}
\label{fig:DR065B_SQUID}
\end{figure}

Single crystals of CeAuSb$_2$ were grown by combining high purity elements (+99.99\% Ce from Ames Lab, +99.9\% Au, +99.99\% Sb Alfa) in a CeAu$_6$Sb$_{12}$ ratio (as determined in Ref.~ \onlinecite{Zhao2016}). The elements were placed in a 5~ml fritted crucible set \cite{Canfield2016} and sealed in an amorphous silica tube with silica wool below and above the set \cite{Canfield2019}.  The growth ampoule was then heated to 1100$^{\circ}$C over 5 hours, dwelled at 1100$^{\circ}$C for 10 hours, cooled to 700$^{\circ}$C over 150 hours. The melt was then decanted in a centrifuge to separate the excess liquid from the CeAuSb$_2$ single crystals \cite{Canfield2019}. Single crystals formed as plate-like crystals (see Fig.~\ref{fig:DR065B_SQUID}) that can readily have dimensions of $5 \times 5 \times 1$~mm$^3$ or larger (with the shortest dimension along the $c$-axis).  Temperature dependent electrical resistivity measurements allowed for evaluation of the residual resistivity ratio, $RRR\sim  6$, consistent with Ref.~\onlinecite{Zhao2016} and DC SQUID magnetisation measurements show the transition into the SDW-A phase occurs at $T_\mathrm{N}\approx 6.5$~K.
 
\subsection{Neutron Diffraction under uniaxial stress}

The sample was mounted in a piezoelectric actuated uniaxial stress apparatus adapted for use in neutron scattering and muon spin rotation experiments. The apparatus and mounting procedure are described in detail in Ref.~\onlinecite{Ghosh2020}. The sample was cut into a bar of approximate dimensions 3.1~mm x  1.3~mm in the $ab$ plane using a wire saw, and polished to a thickness of 0.21~mm along the $c$-axis. 
The sample ends were epoxied into the apparatus using de-gassed Stycast 2850FT prepared with the catalyst 23LV. Cadmium foil (which is strongly neutron absorbing) was used to mask the parts of the apparatus, epoxy exposed to the neutron beam and the end portions of the sample where the strain inhomogeneity was expected to be greatest. As will be discussed below, we observe that some portion of the sample exposed to the neutron beam remained unstrained. The exposed length of sample was approximately 2.2~mm.

The sample holder incorporates a force sensor, and the cell a displacement sensor. Over the entire range of force explored here, the displacement was linear in applied force, indicating that both the sample and epoxy holding it were within their elastic limits. The force sensor was calibrated by hanging weights from the holder, and yields the stress in the sample. Due to deformation of the epoxy holding the sample the displacement sensor cannot be used as an accurate sensor of the strain in the sample.

Neutron diffraction measurements were performed on WISH a time-of-flight (TOF) diffractometer at the ISIS neutron source, UK\cite{Chapon2011}.  
The axis of the uniaxial stress cell was vertical, perpendicular to the (horizontal) scattering plane $(H,0,L)$.  The sample was cooled to base temperature, 1.7~K, (in the presence of exchange gas) before stress was applied. Data were taken at two temperatures (1.7~K and 5.5~K) and a maximum compressive stress of $\sigma_{010} = 440$~MPa. For all stresses, the force and displacement measured by the strain gauge circuits were proportional indicating the sample did not crack. 

The lattice strain along $a$ and $c$ (i.e. transverse to the compressive uniaxial stress) could be resolved in the $d$-spacing of the nuclear Bragg peaks. At non-zero stress, Bragg peaks measured in backscattering geometry (where the $d$-spacing resolution is highest) exhibit a shoulder at the $d$-spacing of the unstrained peak indicating that some of the sample ($\approx20$\%) was unstrained.

Due to limited detector coverage out of the horizontal plane it was not possible to measure the $b$ lattice parameter directly and the elastic constants of CeAuSb$_2$ have not been measured (so $b$ cannot be inferred from the measured strain along the $a$ and $c$ axes). Further details can be found in Appendix \ref{app:nuclear_strain}.

For determination of the UB matrices used to index reflections in the remainder of this work, we assume that the unit cell volume is conserved under uniaxial stress (isochoric distortion). This condition sets the $b$-axis Young's modulus to be 95~GPa, and can be considered to set a lower bound on the magnitude of the $b$-axis strain. 
Comparison of the data here with the phase diagram of Ref.~\onlinecite{Park2018} indicates a Young's modulus of $\approx 60$~GPa. It will be shown that the uncertainty on the $b$ lattice parameter does not impact the conclusions of this paper.

\begin{figure*}[!th]
\centering
\includegraphics[width=0.99\linewidth]{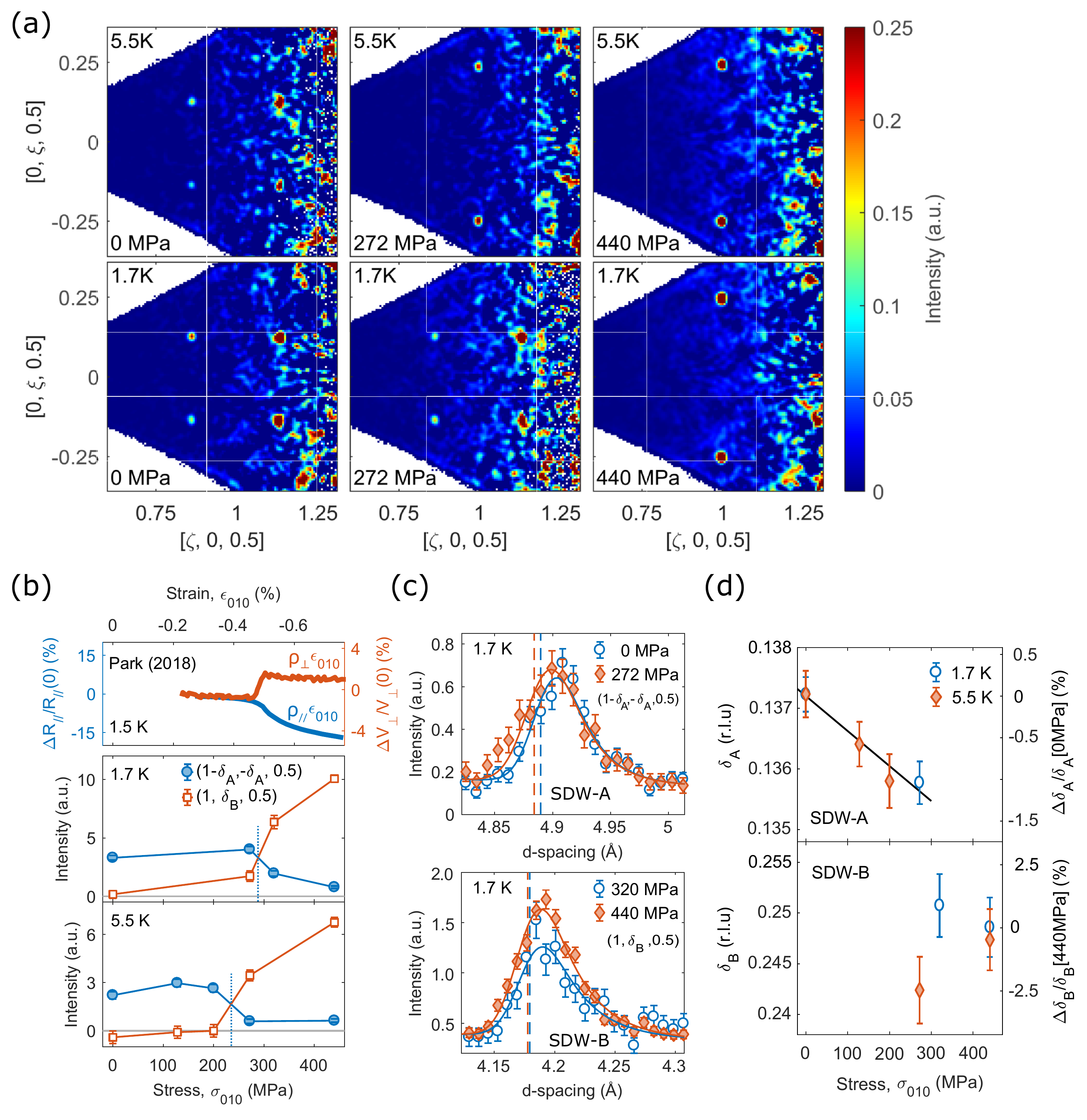}
\caption{(a) Neutron diffraction data in the (H,K,0.5) plane for three values of compressive stress along $[010]$, $\sigma_{010}$, at two temperatures (5.5~K and 1.7~K). The data are smoothed and a planar background has been subtracted. It can be seen that compressive uniaxial stress along $[010]$ induces a mono-domain of single-$\mathbf{q}$ order with a modulation vector rotated by 45$^\circ$ with respect to the zero stress phase. (b) (top) Resistivity for a current applied parallel and perpendicular to the compressive strain as a function of strain at 1.5~K reproduced from Ref.~\onlinecite{Park2018}. (middle, bottom) Integrated intensity of magnetic Bragg peaks in the A \& B phase as a function of stress (after background subtraction) for $T$=1.7, 5.5~K. Vertical blue lines represent the approximate phase boundary determined from the crossing of the A phase and B phase curves.
(c) Magnetic Bragg peaks in the A phase (top) and B phase (bottom) at the minimum and maximum stress at which data were collected at 1.7~K. Solid lines are fits to the peak with a Gaussian convoluted with a back to back exponential, the fitted lattice parameter is indicated by dashed line (which does not coincide with the peak maximum). (d) Refined value of $\delta_A$ (top) and $\delta_B$ (bottom) at two temperatures 1.7~K and 5.5~K as a function of compressive stress along $b$. Solid line is a linear fit. }
\label{fig:5}
\end{figure*}

The single crystal data were integrated using the Mantid software\cite{Mantid}. The data were corrected for the Lorentz factor, normalised to the total current and incident flux and to a vanadium standard sample run to account for detector efficiency. The data were also corrected for absorption using a cylindrical shaped crystal as approximation for the sample shape. The refinements of the nuclear and magnetic structures were performed using the Jana2006 software \cite{Petvrivcek2014} for data collected on a single crystal orientation at 1.7K at two stresses $\sigma_{010}=0$~MPa and $\sigma_{010}=$440~MPa. Structures are displayed using the MVISUALIZE software\cite{MVISUALIZE,Perez-Mato2015}. Cif and mCif files of all the structures refined are reported as supplementary information.\cite{SI}

\section{Results}

\subsection{Stress-induced SDW order (B phase)}
\label{ssec:DiscoveryBphase}

Fig.~\ref{fig:5}(a) shows the scattering intensity in the $(H,K,0.5)$ plane at selected stresses at the two temperatures measured. At zero stress and zero magnetic field, CeAuSb$_2$ exhibits incommensurate SDW order (A phase). The parent $P4/nmm$ space group gives rise to two symmetrically equivalent domains of single-$\mathbf{q}$ order with propagation vectors $\mathbf{q}_{1,2} = (\delta_A,\pm\delta_A,0.5)$, producing four satellite peaks shown in Fig.~\ref{fig:5}a (as seen previously in Ref.~\onlinecite{Marcus2018}).

The compressive stress along the $b$-axis induces a new mono-domain of single-$\mathbf{q}$ SDW order in the B phase with wave-vector $\mathbf{q}_B = (0,\delta_B,0.5)$. The SDW order in A \& B phases doubles the unit cell along the $c$-axis, in contrast to the SDW order in the magnetic-field-induced A$^\prime$ phase \cite{Marcus2018}. In addition it can be seen that the boundary between the A \& B phases occurs at a lower stress at 5.5~K (see Fig.\;\ref{fig:CeAuSb2_PhaseDiag}) and that there is some phase coexistence near the phase boundary (most evident in the 1.7~K data at $\sigma_{010} = 272$~MPa). Phase coexistence would be expected at a first-order transition, but there may also be a contribution from the inhomogeneity of the applied strain.

Fig.~\ref{fig:5}b shows the integrated intensity of magnetic Bragg peaks in the A \& B phases as a function of compressive stress at the two temperatures measured 1.7~K and 5.5~K (after subtraction of a temperature and stress independent linear background).
The resistivity for a current applied parallel and perpendicular to the compressive strain as a function of strain at 1.5~K from Park \textit{et al.} \cite{Park2018} is also reproduced with the axis scaled to the stress used in this experiment using a Young's modulus of $E=60$~GPa, that provides a rough agreement between the observed phase boundary at low temperature.

From Fig.~\ref{fig:5}b it can be seen that at both temperatures measured the SDW intensity in the A phase does not respond strongly to compressive stress along $b$ until the onset stress for the B phase is reached (this can also be seen in the raw data shown in Fig.~\ref{fig:5}c). The onset of the B phase peak intensity is correlated with the reduction in A phase peak intensity, which plateaus at a low value slightly above the background at a stress of $\sigma_{010} \approx 272$~MPa and $\sigma_{010} \gtrsim 320$~MPa at 5.5~K and 1.7~K respectively. The residual A phase peak intensity at the largest stress measured is roughly 20\% of the average zero stress intensity at both temperatures, which is consistent with the volume fraction of the sample under stress determined from the fits to (003) peak (see Appendix~\ref{app:nuclear_strain}).

\subsection{Stress dependence of the SDW modulation}
\label{sec:StressDep}

In Fig.~\ref{fig:5}c we show the scattered intensity for magnetic Bragg peaks in the A and B phases. Data are shown at the minimum and maximum stress measured in each phase at 1.7~K. The peaks were fitted with a Gaussian peak convoluted with a back-to-back exponential (the same profile used to fit the nuclear Bragg peaks). The $d$-spacing of the A phase reflection (indicated by the dashed line in Fig.~\ref{fig:5}c - which does not coincide with the maximum of the peak due to the asymmetric pulse shape from the moderator) exhibits a weak dependence on stress, with a barely resolvable shift to lower $d$-spacing for $\sigma_{010}= 272$~MPa. If the incommensurability of the A phase wavevector, $\delta_A$, were to remain at the zero stress value, the $d$-spacing would be expected to increase sightly by approximately the same magnitude as the observed peak shift (the momentum transfer has only a small component along the compressed axis, $b$, and the $a$ and $c$ lattice parameters are increasing due to the Poisson's ratio of the material). 
There is no resolvable shift is observed in the $d$-spacing of the B phase peak.

Under the assumption of an isochoric lattice distortion, we have determined the incommensurability, $\delta_{A/B}$, of the modulation vectors from the $d$-spacing of the magnetic reflections using a total of 8 peaks (4 symmetrically inequivalent) and 6 peaks (3 symmetrically inequivalent) for the A and B phase respectively. Fig.~\ref{fig:5}d shows the refined incommensurability as a function of stress. Note for the A phase at finite stress the modulation has been assumed to maintain the tetragonal symmetry (i.e. the component of the modulation along $a$ and $b$ has been assumed to be equal) - this is reasonable given the distortion is small and the phase does not appear to couple strongly to stress.

The incommensurability of the A phase at zero stress, $\delta_A=0.137(2)$, is consistent with the value found by Marcus \textit{et al.} \cite{Marcus2018}. These data suggest that $\delta_A$ may decreases slightly with $\sigma_{010}$ (of the order of 1\% over the extent of the phase). The modulation in the B phase at 1.7~K and $\sigma_{010} =440$~MPa stress is $\delta_B\approx0.248(4)$ - it is consistent with a commensurate value $\delta_B=0.25$ for all stresses measured. 
The absolute magnitude of $\delta_B$ (and to a lesser extent $\delta_A$) will of course depend on the $b$ lattice parameter: in the refinement we have assumed an isochoric distortion, however the effect of assuming the material is softer with a Young's modulus of $E\approx60$~GPa (determined from the approximate phase boundary at 1.7~K) is an order of magnitude smaller than the uncertainty on $\delta_B$ at $\sigma_{010} =440$~MPa.

\subsection{Refinement of the single crystal data}

\label{Sec:Refinement}

\subsubsection{Zero strain phase (A phase)}

\begin{figure}[!htb]
\includegraphics[width=1\linewidth]{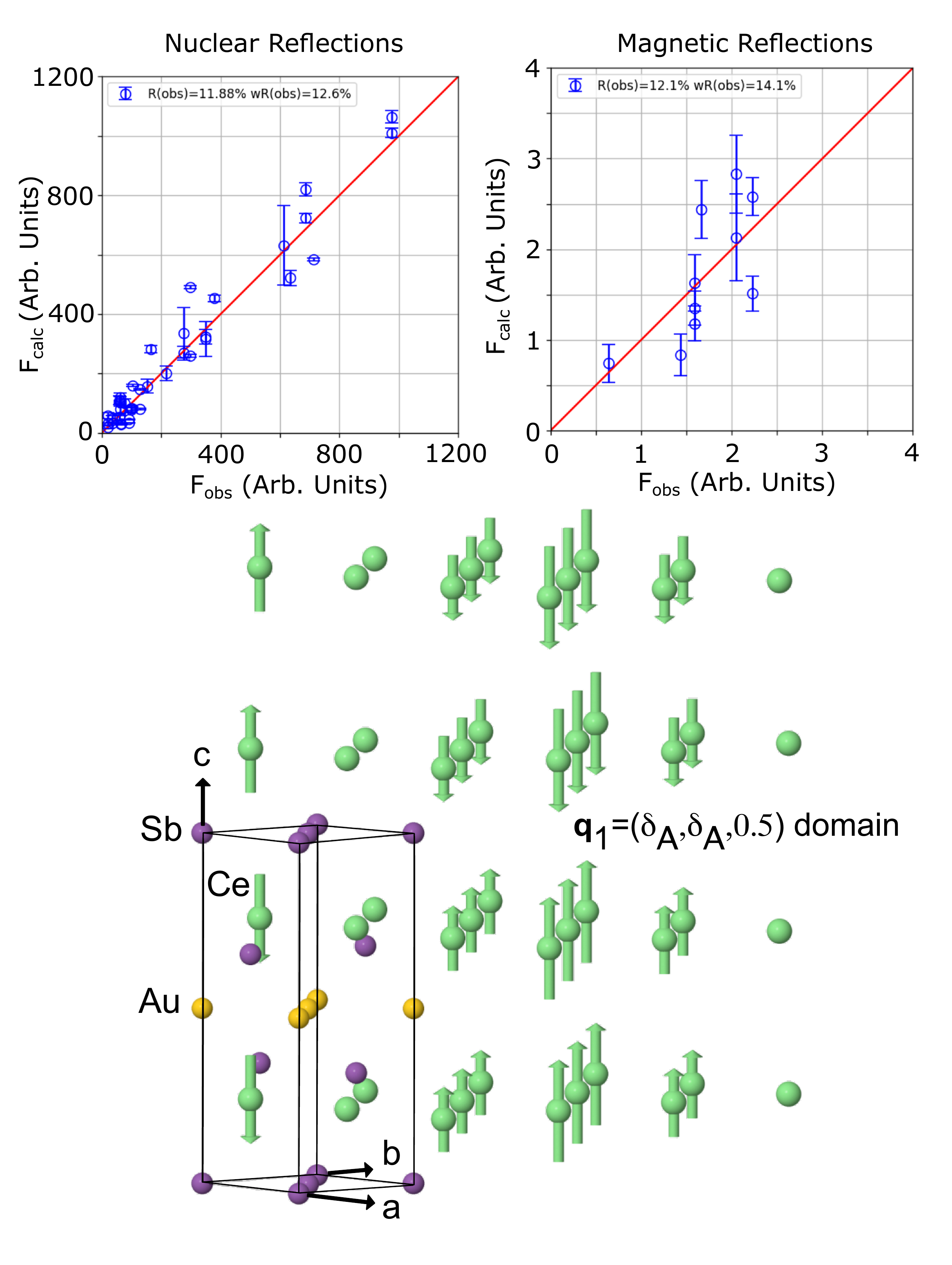}
\caption{A phase structure. (Top) Observed and calculated intensities from the refinement of the nuclear and magnetic structure at 0~MPa and 1.7~K using Jana2006. The nuclear structure refinement has $R=11.9\%$ and $R_w=12.6\%$ and the magnetic structure refinement has $R=12.1\%$ and $R_w=14.1\%$. (Bottom) drawing of the nuclear and magnetic structure at 0~MPa and 1.7~K (only cerium atoms shown in magnetic structure for clarity). The magnetic structure shows a single $\mathbf{q}_{1}=(\delta_A,\delta_A,0.5)$ domain.}
\label{fig:Zero}
\end{figure}

The zero stress data were refined from the published structure with the tetragonal space group $P4/nmm$ \cite{Sologub1994}. A total of 41 viable nuclear reflections were integrated over an ellipsoid in $\mathbf{Q}$-space with axes determined for each reflection from the estimated covariance of the data in a sphere of radius governed by the TOF width of the peak. To reduce the number of refinable parameters a single isotropic thermal factor has been defined for all atomic species. The extinction correction used in the refinement is a type-II Becker and Coppens isotropic model \cite{Becker1974}. The results of the nuclear structure refinement are shown in Table.~\ref{tab:ZeroStr}, the agreement between calculated and observed structure factor and a sketch of the structure are reported in Fig.~\ref{fig:Zero}.

\renewcommand{\tabcolsep}{15pt} 
\renewcommand\arraystretch{1.25} 
\begin{table}[]
\centering
\begin{tabular}{ll}
\hline
$T$ = 1.7~K & \multicolumn{1}{c}{0~MPa} \\ \hline
Spacegroup & $P4/nmm$\\ \hline
$a$(\AA) & 4.3976(3)  \\
$b$(\AA) & 4.3976(3)  \\
$c$(\AA) & 10.30644(9)  \\ \hline
Ce,$z$ & 0.7545(14)  \\
Sb2,$z$ & 0.3261(14)  \\
U$_{\mathrm{iso}}$ & 0.0047(18) \\ \hline
Reflections & 41  \\
$R$ & 11.88\% \\
$R_w$ & 12.61\% \\ \hline
\end{tabular}
\caption{Lattice parameters and nuclear structure refinement results for data at 0~MPa at 1.7~K.  Atomic positions are in fractional coordinates: Ce 2c$(0,0.5,z)$, Au 2b$(-0.5,0.5,0.5)$, Sb1 2a$(0,0,0)$, Sb2 2c$(0,0.5,z)$. An isotropic thermal displacement parameter $U_{\mathrm{iso}}$ was constrained to be the same for all atoms.}
\label{tab:ZeroStr}
\end{table}

The magnetic superspace group for the SDW A phase has been determined with the help of group theoretical calculation using the ISODISTORT and ISOTROPY software\cite{Hatch2003, stokes2016}. Two irreducible representation (irreps) mS2 and mS4, corresponding to the superspace groups $Cmme1^{\prime}(0\beta1/2)s00s$ and $Cmme1^{\prime}(0\beta1/2)s0ss$ respectively, have been found to refine the data with the same reliability parameters. The difference between the two irreps regards the relative phase of the Ce moment at different $z$-coordinate, being antiferromagnetic for mS2 and ferromagnetic for mS4. The refined SDW has an amplitude of $0.66(4)\mu_B$. The refinement has been conducted on 10 magnetic reflections taking into account both magnetic domains and by fixing the domain fraction to 0.5. The reliability factor for the magnetic refinement are $R=12.1\%$ and $R_w=14.1\%$. The value of the SDW amplitude in zero strain is significantly smaller than what observed by Marcus \textit{et al.} \cite{Marcus2018}. The reason for this discrepancy is still not clear. The final agreement between the observed and calculated structure factors is shown in Fig.~\ref{fig:Zero}. As mentioned earlier, the SDW in the A phase has single-$\mathbf{q}$ domains \cite{Marcus2018}, a $\mathbf{q}_1$ domain is shown in Fig.~\ref{fig:Zero}.

\subsubsection{High strain phase (B phase)}
\label{Sec:High_strain}
\begin{figure}[!hbt]
\includegraphics[width=0.99\linewidth]{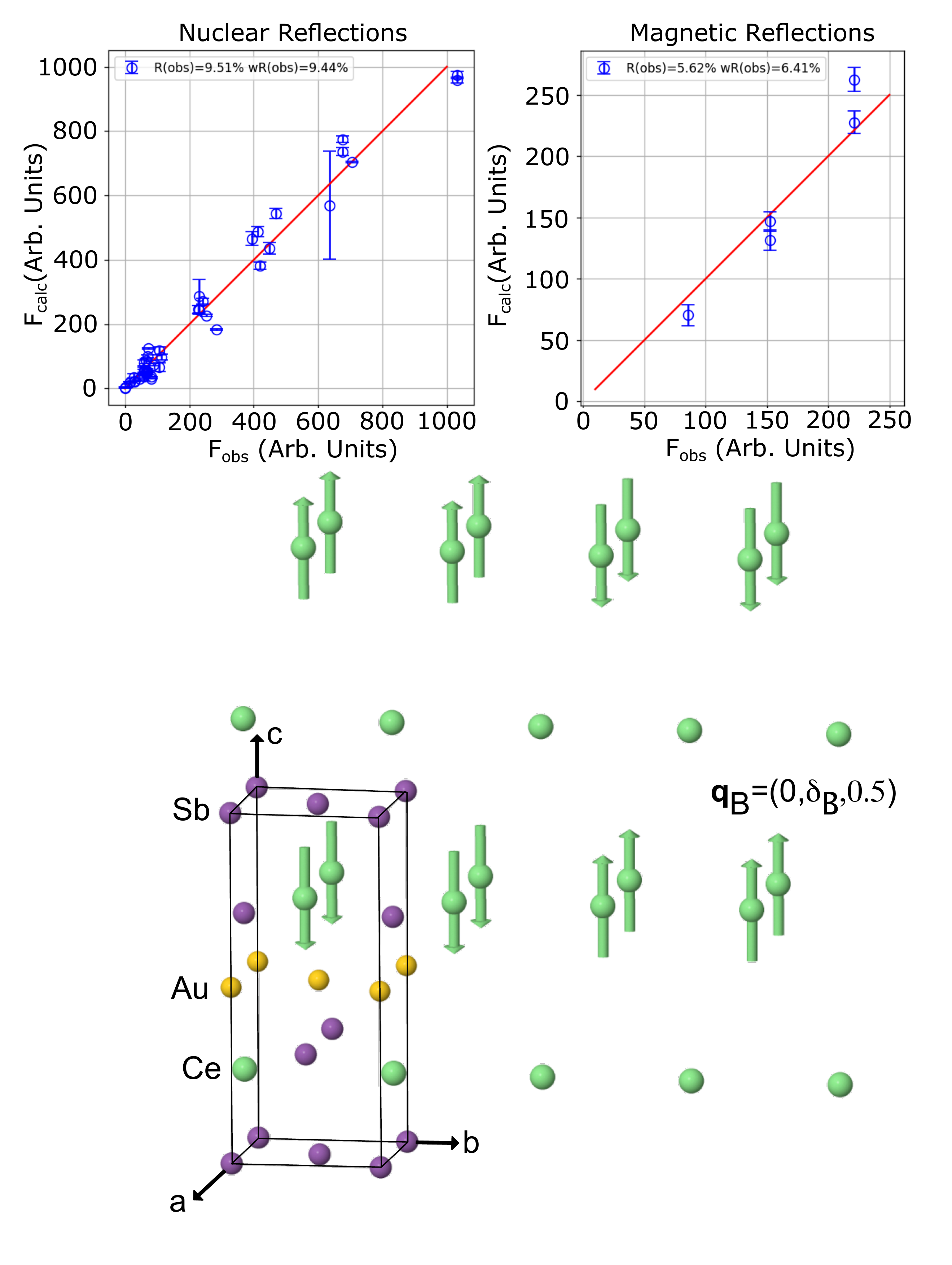}
\caption{B phase structure. (Top) Observed and calculated intensities from the refinement of the nuclear and magnetic structure at 440~MPa and 1.7~K using Jana2006. The nuclear structure refinement has $R=9.5\%$ and $R_w=9.4\%$ and the magnetic structure refinement has $R=5.6\%$ and $R_w=6.4\%$ (commensurate $A_bem2$ structure).  (Bottom) drawing of the nuclear and magnetic structure at 440~MPa and 1.7~K (only cerium atoms shown in magnetic structure for clarity). }
\label{fig:high}
\end{figure}

The single crystal data at finite stress were refined with the orthorhombic space group $Pmmn$ for the nuclear structure, derived from the action of the orthorhombic $\Gamma_2^+$ strain on the $P4/nmm$ space group. As for the zero strain structure  a single isotropic thermal parameter has been refined and an isotropic Becker and Coppens \cite{Becker1974} Type-II model has been used for extinction correction. The structure parameters and reliability factors are reported in Table.~\ref{tab:HighStr}, the agreement between observed and calculated structure factors and a sketch of the nuclear structure are reported in Fig.~\ref{fig:high}

\renewcommand{\tabcolsep}{15pt} 
\renewcommand\arraystretch{1.25} 
\begin{table}[]
\centering
\begin{tabular}{ll}
\hline
$T$ = 1.7~K & {440~MPa} \\ \hline
Spacegroup &  $Pmmn$ \\ \hline
$a$(\AA)  & 4.4035(3) \\
$b$(\AA)  & 4.3772 \\
$c$(\AA)  & 10.3404(1) \\ \hline
Ce,$z$ &  0.7652(11) \\
Au,$z$ &  0.4950(8) \\
Sb1,$z$ & 0.0026(11) \\
Sb2,$z$ & 0.3192(10) \\
U$_{\mathrm{iso}}$ & {0.0054(13)} \\ \hline
Reflections & 41 \\
$R$  & 9.51 \\
$R_w$ & 9.44 \\ \hline
\end{tabular}
\caption{Lattice parameters and nuclear structure refinement results for data at 440~MPa compressive stress along the $b$-axis at 1.7~K. The $a/c$ lattice parameters were determined from fits to nuclear Bragg peaks, the $b$ lattice parameter assumes an isochoric distortion. Atomic positions are in fractional coordinates: Ce 2c$(0,0.5,z)$, Au 2b$(0,0,z)$, Sb1 2a$(0,0,z)$, Sb2 2c$(0,0.5,z)$. An isotropic thermal displacement parameter $U_{\mathrm{iso}}$ was constrained to be the same for all atoms. The $a$ and $c$ lattice parameters are increased with compressive strain because of the Poisson effect.}
\label{tab:HighStr}
\end{table}

The experimentally determined propagation vector in the B phase, $\mathbf{q}_B=(0,\delta_B,1/2)$, is commensurate with $\delta_B=1/4$ within the accuracy of our measurements, as discussed Sec.\;\ref{sec:StressDep}. Furthermore there is a symmetry reason to lock the propagation vector to the commensurate value. The latter comes from the fact that when $\delta_B=1/4$, the Landau free energy decomposition allows the presence of additional “lock-in” terms, $h_j^8+h_j^{*8}$, where ($h_j,h_j^*$) are the complex order parameters transformed by $mB_j(j=1-4)$ irreducible representations (irreps) of the $Pmmn$ space group, associated with the $\mathbf{q}_B$ propagation vector. The invariance of these terms can be verified using the matrix operators for the generating symmetry elements summarised in Table ~\ref{tab:Sym} in the Appendix~\ref{app:lockin}. It worth to point out that the lock-in commensurate phases have been also observed in some other Ce-based intermetallic systems such as CeIrGe$_3$\cite{Anand2018} and CeRhGe$_3$ \cite{Hillier2012}. Taking into account these facts, we approached the refinement of the magnetic structure in the B phase of CeAuSb$_2$ assuming a commensurate propagation vector.
In this scenario, the symmetry of the magnetic structure depends on the global phase of the modulation, resulting in a several possible origin choices. Another important symmetry aspect is the existence of linear-cubic free energy invariants which provide a coupling between $mB_1(\eta_1,\eta_1^*)$ and $mB_3(\eta_3,\eta_3^*)$, $\eta_1\eta_3^3+\eta_1^*\eta_3^{*3}$, and between $mB_2(\eta_2,\eta_2^*)$ and $mB_4(\eta_4,\eta_4^*)$, $\eta_2\eta_4^3+\eta_2^*\eta_4^{*3}$, order parameters. These invariants are symmetric with respect to the subscript indices and the coupling terms $\eta_3\eta_1^3+\eta_3^*\eta_1^{*3}$ and $\eta_4\eta_2^3+\eta_4^*\eta_2^{*3}$ are also allowed. This implies that in the case of commensurate ordering, the corresponding order parameters can be mixed without changing the magnetic symmetry of the system. 
The $mB_1$ and $mB_3$ irreps transform the magnetic modes with the moments along the $a$-axis of the $Pmmn$ space group, while $mB_2$ and $mB_4$ transform the modes with the moments along the $b$- and $c$-axes. The magnetic phases of the spin density waves localized on the Ce sites with z=0.235 and z=0.765 in the parent structure, differ by $\pi$ in the magnetic modes which belong to the different irreps. Quantitative refinement of the measured magnetic intensities revealed that, similar to the A phase, the Ce moments in the B phase are predominantly polarized along the $c$-axis which is common for both tetragonal $P4/nmm$ and $Pmmn$ space groups.  Moreover, the refinement has been found to be sensitive to the admixture between $mB_2$ and $mB_4$ irreps, yielding equal weight for both order parameters, as well as to their relative magnetic phases. The best fitting quality was achieved in the model shown in Fig.~\ref{fig:high} with reliability factors $R=5.62\%$ and $R_W=6.41\%$. This solution implies the orthorhombic polar symmetry $A_bem2$, making CeAuSb$_2$ an exciting system, where loss of spatial inversion is driven by magnetic ordering. This phenomenon is well known for insulators such as type II multiferroics\cite{Mostovoy2006}, but it rather rare for metals \cite{Princep2020,Feng2021}.

Refinement of the neutron diffraction data assuming the incommensurate propagation vector provided a worse fitting quality with a $R_W$ reliability factor of 8.6\% compared to 6.4\% for the commensurate $A_bem2$ with the same number of refinable variables. In this case, two magnetic superspace groups $Pmmm1^{\prime}(0,\beta,1/2)s0ss$ and $Pmmm1^{\prime}(0,\beta,1/2)s00s$ associated with $mB_2$ and $mB_4$ irreps cannot be distinguished. In both models, the Ce moments are aligned along the $c$-axis and the difference stands to the relative magnetic phases for the spin density waves localized on the Ce sites with different $z$-coordinate in the parent structure. No admixture of the irreps is allowed for the incommensurate propagation vector. Due to the worse fitting quality, the observed propagation vector and the symmetry reasons discussed above, we believe the incommensurate scenario is less likely and we do not discuss it further.
The proposed commensurate magnetic structure, shown in Fig.~\ref{fig:high}, involves alternation of the Ce layers which carry magnetic moments of $1.15(4)\mu_B$ at 1.7K with zero-moment layers. In the former, the ordering is up, up, down, down upon propagation along the $b$-axis.

\section{Discussion}
\label{sec:discussion}

\subsection{The magnetism of CeAuSb$_2$}

There are two paradigms in which to consider the magnetism in this compound: the localised limit and the itinerant limit in which the Ce moments are screened by the conduction electrons. As discussed above, previous neutron scattering measurements suggest that the itinerant limit is appropriate here \cite{Marcus2018}. The refined wavevector of the B phase is consistent within the error with a commensurate modulation, $\mathbf{q}_B = (0, 0.25, 0.5)$. We note that this conclusion is not impacted by the negligible systematic uncertainty on the $b$-axis lattice parameter. A commensurate wavevector does not necessarily imply local-moment order. For example, dilute chromium alloys\cite{Fawcett1988} and the iron pnictide superconductor BaFe$_2$As$_2$\cite{Pratt2011} exhibit a first-order incommensurate-commensurate transition that arises from slightly imperfect nesting between electron and hole-pockets, which in both of the above compounds can be tuned with doping. 
Unlike an incommensurate sinusoidal modulation, the free energy of a commensurate modulation depends on the phase. It might be energetically favourable to preserve a constant moment across all sites (which would hint at local moment behaviour), or conversely to have nodes of zero amplitude on specific sites (which is incompatible with local moments). This can force a modulation to `lock-in' on a commensurate value even if the ideal nesting is slightly incommensurate\cite{Lee1993}.
For example, such a mechanism is believed to be behind the incommensurate-commensurate transitions of the charge-density wave in 2H-TaSe$_2$\cite{Fleming1984}.

Overall the observed magnetic structure of the B phase that includes Ce sites with zero magnetic moment, shown in Fig.~\ref{fig:high}, and the detailed symmetry analysis reported in Sec. III-C suggest that the order in CeAuSb$_2$ is closer to the itinerant limit.  We note that the presence of sites with zero-magnetic moments is not unique in metallic systems and it has been observed in other SDW ordered materials such as in the C$_4$ phase of iron-based superconductors\cite{Avci2014, Allred2016_ATBK}

The present work taken together with the previous neutron study \cite{Marcus2018} shows that the SDW state of CeAuSb$_2$ may be switched by the application of magnetic field or uniaxial stress. However, it should be noted that magnetic field and uniaxial stress couple to the SDW in different ways. The field causes a field-dependent exchange splitting between up- and down-spin bands, whereas the uniaxial stress modifies the hybridisation, for example, between the $f$ electrons and conduction electrons.   At the lowest temperature the SDW phase transitions are first order with a change in the symmetry of the order parameter. The ordering in SDW systems can be often be understood in terms of the wavevector-dependent susceptibility $\chi(\mathbf{q})$ calculated from the Lindhard function. In the case of CeAuSb$_2$ we would then expect $\chi(\mathbf{q})$ to be sufficiently field and stress dependent that the SDW ordering wavevector can switch between different $\mathbf{q}$'s. To date, a detailed connection between the electronic structure/Fermi surface the ordering wavevectors has not been identified, however, published band structures\cite{Marcus2018,Yumnam2019} do offer possibilities for nesting. We note that a Fermi surface that shows topological reconstruction as a function of magnetic field has been proposed in the sister compound Sr$_3$Ru$_2$O$_7$ which also exhibits a SDW controlled by magnetic field \cite{Efremov2018,Lester2015}.

\subsection{Transport signature of the B phase}

Previous transport measurements have mapped out the temperature-strain-magnetic field phase diagram of CeAuSb$_2$ \cite{Zhao2016,Park2018,Park2018a}. The neutron diffraction data presented here are qualitatively consistent with the published phase diagram \cite{Park2018,Park2018a}. In particular we observe the onset of magnetic Bragg scattering associated with the B phase at a lower stress at 5.5~K than 1.7~K. 
The present data shed new light on the anisotropic transport observed inside the B phase. The onset of B phase order as a function of stress at low temperature is associated with a first-order step in the resistivity perpendicular to the axis of compression\cite{Park2018}.  We can now identify the axis of the enhanced resistivity as perpendicular to the observed wavevector in the B phase (i.e. $\rho_\perp > \rho_{\|}$). The opposite trend is observed in Sr$_3$Ru$_2$O$_7$\cite{Borzi2007,Lester2015} and in Cr \cite{Muir1971} ($\rho_{\|} > \rho_\perp$). As the B phase exhibits only one domain the enhanced resistivity is intrinsic to the order (i.e. not due to scattering at domain walls).

SDW order causes Fermi surface reconstruction \cite{Millis2007} and therefore changes in transport properties such as resistivity ($\rho$) and Hall number ($n_H$) in materials such as Cr \cite{Fawcett1988} and Sr$_3$Ru$_2$O$_7$ \cite{Lester2015}. Here $\rho$ and $n_H$ derive from integrals of the Fermi velocity and other quantities over the Fermi surface.  We argue, by analogy, that this mechanism is responsible for dramatic changes seen in $\rho$ for CeAuSb$_2$ on entering the $B$-phase [See Fig.~~\ref{fig:5}(b)]. The re-constructed Fermi surface produced by the single-$\mathbf{q}$ SDW leads to an anisotropic $\rho$ as is the case for Cr and Sr$_3$Ru$_2$O$_7$.

\section{Conclusion}

The main result of this work is that compressive uniaxial stress applied along the [010] axis of tetragonal CeAuSb$_2$ induces a phase (B phase) with a mono-domain of single-$\mathbf{q}$ SDW order with commensurate wavevector $\mathbf{q}_B=(0, 0.25, 0.5)$ producing a structure that breaks spatial inversion symmetry. The B phase of CeAuSb$_2$ is one of small number of metals \cite{Princep2020,Feng2021} where the loss of spatial inversion is driven by magnetic ordering.
The component of $\mathbf{q}_B$ in the $ab$-plane is parallel to the direction of the applied stress. This contrasts with A phase (in the absence of stress) where the SDW has wavevectors $\mathbf{q}_A=(0.136,\pm 0.136,0.5)$. We believe that the change in SDW wavevector leads to a Fermi surface reconstruction which is reflected in a change in transport anisotropy. CeAuSb$_2$ is a good system to test this mechanism because a single domain SDW can be prepared. A similar mechanism may occur in other materials where magnetic field or strain-dependent anomalies in resistivity are observed including Sr$_3$Ru$_2$O$_7$\cite{Borzi2007,Lester2015}, URu$_2$Si$_2$\cite{Knafo2016}, CeNiGe$_3$~\cite{Mun2010},CeRh$_2$Si$_2$~\cite{Knafo2010} and  YbNiSi$_3$~\cite{Bud2007}.

\begin{acknowledgments}
We acknowledge funding and support from the Engineering and Physical Sciences Research Council (EPSRC) Centre for Doctoral Training in Condensed Matter Physics (CDT-CMP), Grant No. EP/L015544/1. The authors thanks the Science and Technology Facility Council (STFC) for the provision of neutron beam time at ISIS (UK). Work done at Ames Laboratory (PCC, RAR) was supported by the U.S. Department of Energy, Office of Basic Energy Science, Division of Materials Sciences and Engineering. Ames Laboratory is operated for the U.S. Department of Energy by Iowa State University under Contract No. DE-AC02-07CH11358.  RAR was also supported by the Gordon and Betty Moore Foundation’s EPiQS Initiative through Grant GBMF4411.
\end{acknowledgments}

\appendix

\renewcommand{\tabcolsep}{15pt} 
\renewcommand\arraystretch{1.25} 
\begin{table*}[]
\centering
\begin{tabular}{|l|l|l|l|l|}
\hline
  &$mB_1(\eta_1,\eta_1^*)$ & $mB_2(\eta_2,\eta_2^*)$ & $mB_3(\eta_3,\eta_3^*)$ & $mB_4(\eta_4,\eta_4^*)$ \\ \hline
& & & & \\	
$\{2_z|1/2,1/2,0\}$  
& $\left( \begin{array}{cc} 0 & e^{\frac{1}{4} \pi i} \\ e^{-\frac{1}{4} \pi i} & 0 \\ \end{array} \right)$ 
& $\left( \begin{array}{cc} 0 & e^{\frac{5}{4} \pi i} \\ e^{\frac{3}{4} \pi i} & 0 \\ \end{array} \right)$ 
& $\left( \begin{array}{cc} 0 & e^{\frac{5}{4} \pi i} \\ e^{\frac{3}{4} \pi i} & 0 \\ \end{array} \right)$ 
& $\left( \begin{array}{cc} 0 & e^{\frac{1}{4} \pi i} \\ e^{-\frac{1}{4} \pi i} & 0 \\ \end{array} \right)$ \\ 
& & & & \\
$\{2_y|0,1/2,0\}$  
& $\left( \begin{array}{cc} e^{\frac{1}{4} \pi i} & 0 \\  0 & e^{-\frac{1}{4} \pi i} \\ \end{array} \right)$ 
& $\left( \begin{array}{cc} e^{\frac{1}{4} \pi i} & 0 \\  0 & e^{-\frac{1}{4} \pi i} \\ \end{array} \right)$ 
& $\left( \begin{array}{cc} e^{\frac{5}{4} \pi i} & 0 \\  0 & e^{\frac{3}{4} \pi i} \\ \end{array} \right)$ 
& $\left( \begin{array}{cc} e^{\frac{5}{4} \pi i} & 0 \\  0 & e^{\frac{3}{4} \pi i} \\ \end{array} \right)$ \\
& & & & \\
$\{-1|0,0,0\}$  
& $\left( \begin{array}{cc} 0 & 1 \\  1 & 0 \\ \end{array} \right)$ 
& $\left( \begin{array}{cc} 0 & 1 \\  1 & 0 \\ \end{array} \right)$ 
& $\left( \begin{array}{cc} 0 & 1 \\  1 & 0 \\ \end{array} \right)$ 
& $\left( \begin{array}{cc} 0 & 1 \\  1 & 0 \\ \end{array} \right)$ \\
& & & & \\
$\{1|0,1,0\}$  
& $\left( \begin{array}{cc} e^{\frac{1}{2} \pi i} & 0 \\  0 & e^{-\frac{1}{2} \pi i} \\ \end{array} \right)$ 
& $\left( \begin{array}{cc} e^{\frac{1}{2} \pi i} & 0 \\  0 & e^{-\frac{1}{2} \pi i} \\ \end{array} \right)$ 
& $\left( \begin{array}{cc} e^{\frac{1}{2} \pi i} & 0 \\  0 & e^{-\frac{1}{2} \pi i} \\ \end{array} \right)$ 
& $\left( \begin{array}{cc} e^{\frac{1}{2} \pi i} & 0 \\  0 & e^{-\frac{1}{2} \pi i} \\ \end{array} \right)$ \\
& & & & \\
$\{1|0,0,1\}$  
& $\left( \begin{array}{cc} -1 & 0 \\  0 & -1 \\ \end{array} \right)$ 
& $\left( \begin{array}{cc} -1 & 0 \\  0 & -1 \\ \end{array} \right)$ 
& $\left( \begin{array}{cc} -1 & 0 \\  0 & -1 \\ \end{array} \right)$ 
& $\left( \begin{array}{cc} -1 & 0 \\  0 & -1 \\ \end{array} \right)$\\ 
& & & & \\
$T$  
& $\left( \begin{array}{cc} -1 & 0 \\  0 & -1 \\ \end{array} \right)$ 
& $\left( \begin{array}{cc} -1 & 0 \\  0 & -1 \\ \end{array} \right)$ 
& $\left( \begin{array}{cc} -1 & 0 \\  0 & -1 \\ \end{array} \right)$ 
& $\left( \begin{array}{cc} -1 & 0 \\  0 & -1 \\ \end{array} \right)$\\ 
& & & & \\
\hline
\end{tabular}
\caption{Matrices of the irreducible representations of $Pmmn$ space group, associated with q=(0,1/4,1/2) propagation vector\cite{Aroyo2006}. $T$ is time reversal operator.}
\label{tab:Sym}
\end{table*}

\section{Measurement of the applied strain}
\label{app:nuclear_strain}

\begin{figure*}
\centering
\includegraphics[width=0.9\linewidth]{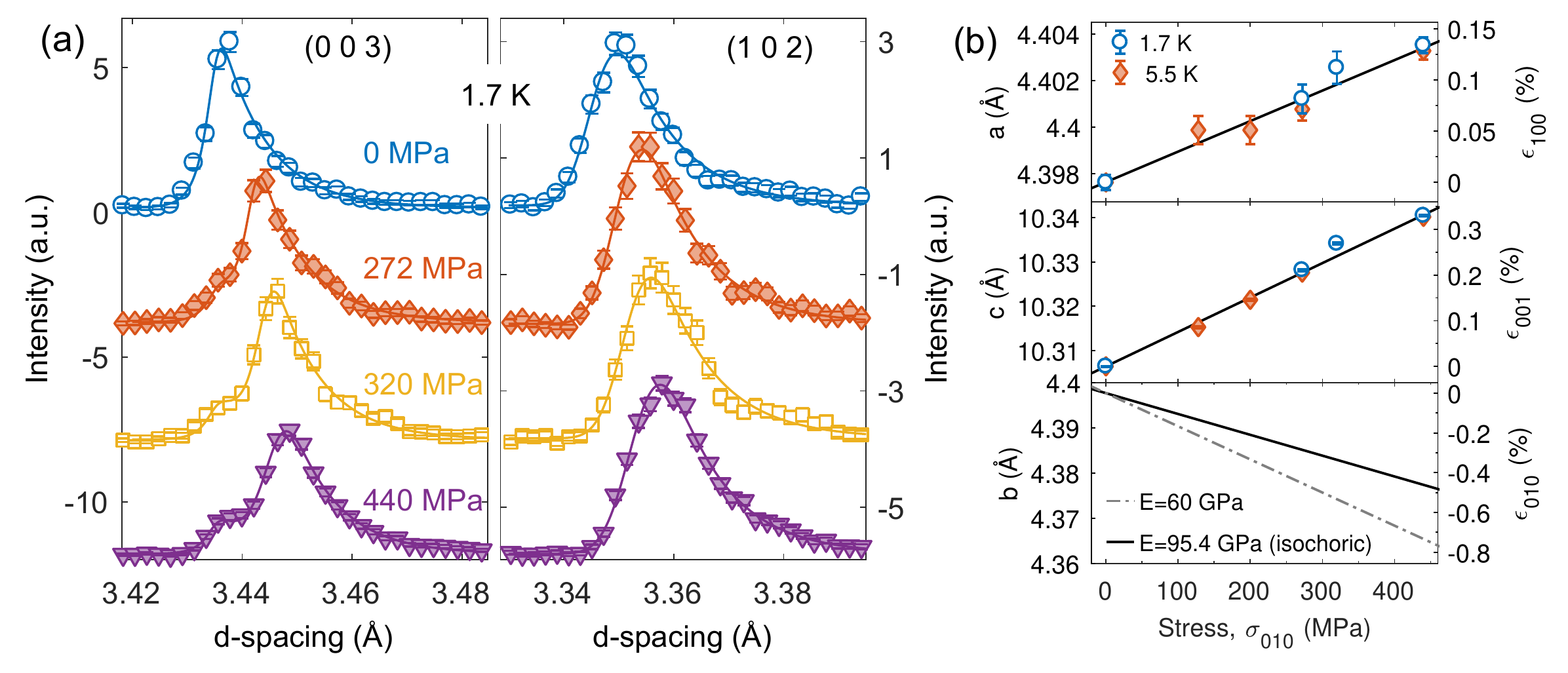} 
\caption{(a) The intensity of (003) and (102) nuclear peaks plotted as a function of $d$-spacing for various values of compressive stress along the $b$-axis (curves are offset for clarity). Solid lines are fits with a back-to-back exponential convoluted with a Gaussian. The (003) peak was measured at higher resolution, at non-zero stress the peak exhibits a shoulder at the zero stress position indicating a portion of the sample ($\approx 20$\%) was not stressed. The peak positions shift due to larger $d$-spacing with increasing compression along the $b$-axis as the $a$- and $c$-axis lattice parameters increase. (b) The  $a$ and $c$ lattice parameters at 1.7 K and 5.5 K as a function of compressive stress along the $b$-axis, $\sigma_{010}$, and the predicted $b$-axis lattice parameter for two values of the Young’s modulus ($b$ could not be measured directly). The lattice parameters were refined from the $d$-spacing of 8 nuclear peaks.
\label{sfig:NucFoc}
}
\end{figure*}

Fig.~\ref{sfig:NucFoc}(a) shows the intensity of (003) and (102) nuclear peaks at a different compressive stresses. Under compressive stress the peaks shift to higher $d$-spacing as the $a$ and $c$ lattice parameters increase due to the Poisson's ratios of the material. The stress dependence of the $a$ and $c$ lattice parameters is shown in Fig.~\ref{sfig:NucFoc}(b)), obtained from a fit of 8 nuclear peaks using a Gaussian peak convoluted with a back-to-back exponential \cite{VonDreele1982}. 

At finite stress several nuclear peaks that were measured in backscattering geometry, where the $d$-spacing resolution is highest, such as the (003) peak shown in Fig.~\ref{sfig:NucFoc}(a). The peaks are resolution limited, even in the strained data.

\section{Lock-in terms}
\label{app:lockin}

Table~\ref{tab:Sym} reports the transformation matrices, defined  in a complex basis, for the generators of the $Pmmn$ space group. The $mB_j(j=1-4)$  irreducible representations are all time odd and the matrices can be used to verify the invariance of the free energy terms discussed in Sec.~\ref{Sec:High_strain}.


\begin{thebibliography}{48}%
\makeatletter
\providecommand \@ifxundefined [1]{%
 \@ifx{#1\undefined}
}%
\providecommand \@ifnum [1]{%
 \ifnum #1\expandafter \@firstoftwo
 \else \expandafter \@secondoftwo
 \fi
}%
\providecommand \@ifx [1]{%
 \ifx #1\expandafter \@firstoftwo
 \else \expandafter \@secondoftwo
 \fi
}%
\providecommand \natexlab [1]{#1}%
\providecommand \enquote  [1]{``#1''}%
\providecommand \bibnamefont  [1]{#1}%
\providecommand \bibfnamefont [1]{#1}%
\providecommand \citenamefont [1]{#1}%
\providecommand \href@noop [0]{\@secondoftwo}%
\providecommand \href [0]{\begingroup \@sanitize@url \@href}%
\providecommand \@href[1]{\@@startlink{#1}\@@href}%
\providecommand \@@href[1]{\endgroup#1\@@endlink}%
\providecommand \@sanitize@url [0]{\catcode `\\12\catcode `\$12\catcode
  `\&12\catcode `\#12\catcode `\^12\catcode `\_12\catcode `\%12\relax}%
\providecommand \@@startlink[1]{}%
\providecommand \@@endlink[0]{}%
\providecommand \url  [0]{\begingroup\@sanitize@url \@url }%
\providecommand \@url [1]{\endgroup\@href {#1}{\urlprefix }}%
\providecommand \urlprefix  [0]{URL }%
\providecommand \Eprint [0]{\href }%
\providecommand \doibase [0]{http://dx.doi.org/}%
\providecommand \selectlanguage [0]{\@gobble}%
\providecommand \bibinfo  [0]{\@secondoftwo}%
\providecommand \bibfield  [0]{\@secondoftwo}%
\providecommand \translation [1]{[#1]}%
\providecommand \BibitemOpen [0]{}%
\providecommand \bibitemStop [0]{}%
\providecommand \bibitemNoStop [0]{.\EOS\space}%
\providecommand \EOS [0]{\spacefactor3000\relax}%
\providecommand \BibitemShut  [1]{\csname bibitem#1\endcsname}%
\let\auto@bib@innerbib\@empty
\bibitem [{\citenamefont {Fisk}\ \emph {et~al.}(1986)\citenamefont {Fisk},
  \citenamefont {Ott}, \citenamefont {Rice},\ and\ \citenamefont
  {Smith}}]{Fisk1986}%
  \BibitemOpen
  \bibfield  {author} {\bibinfo {author} {\bibfnamefont {Z.}~\bibnamefont
  {Fisk}}, \bibinfo {author} {\bibfnamefont {H.~R.}\ \bibnamefont {Ott}},
  \bibinfo {author} {\bibfnamefont {T.~M.}\ \bibnamefont {Rice}}, \ and\
  \bibinfo {author} {\bibfnamefont {J.~L.}\ \bibnamefont {Smith}},\ }\href
  {\doibase 10.1038/320124a0} {\bibfield  {journal} {\bibinfo  {journal}
  {Nature}\ }\textbf {\bibinfo {volume} {320}},\ \bibinfo {pages} {124}
  (\bibinfo {year} {1986})}\BibitemShut {NoStop}%
\bibitem [{\citenamefont {Knebel}\ \emph {et~al.}(2011)\citenamefont {Knebel},
  \citenamefont {Aoki},\ and\ \citenamefont {Flouquet}}]{Knebel2001}%
  \BibitemOpen
  \bibfield  {author} {\bibinfo {author} {\bibfnamefont {G.}~\bibnamefont
  {Knebel}}, \bibinfo {author} {\bibfnamefont {D.}~\bibnamefont {Aoki}}, \ and\
  \bibinfo {author} {\bibfnamefont {J.}~\bibnamefont {Flouquet}},\ }\href
  {\doibase https://doi.org/10.1016/j.crhy.2011.05.002} {\bibfield  {journal}
  {\bibinfo  {journal} {Comptes Rendus Physique}\ }\textbf {\bibinfo {volume}
  {12}},\ \bibinfo {pages} {542} (\bibinfo {year} {2011})},\ \bibinfo {note}
  {superconductivity of strongly correlated systems}\BibitemShut {NoStop}%
\bibitem [{\citenamefont {Doniach}(1977)}]{Doniach1977}%
  \BibitemOpen
  \bibfield  {author} {\bibinfo {author} {\bibfnamefont {S.}~\bibnamefont
  {Doniach}},\ }\href {\doibase https://doi.org/10.1016/0378-4363(77)90190-5}
  {\bibfield  {journal} {\bibinfo  {journal} {Physica B+C}\ }\textbf {\bibinfo
  {volume} {91}},\ \bibinfo {pages} {231} (\bibinfo {year} {1977})}\BibitemShut
  {NoStop}%
\bibitem [{\citenamefont {Shishido}\ \emph {et~al.}(2005)\citenamefont
  {Shishido}, \citenamefont {Settai}, \citenamefont {Harima},\ and\
  \citenamefont {Ōnuki}}]{Shishido2005}%
  \BibitemOpen
  \bibfield  {author} {\bibinfo {author} {\bibfnamefont {H.}~\bibnamefont
  {Shishido}}, \bibinfo {author} {\bibfnamefont {R.}~\bibnamefont {Settai}},
  \bibinfo {author} {\bibfnamefont {H.}~\bibnamefont {Harima}}, \ and\ \bibinfo
  {author} {\bibfnamefont {Y.}~\bibnamefont {Ōnuki}},\ }\href {\doibase
  10.1143/JPSJ.74.1103} {\bibfield  {journal} {\bibinfo  {journal} {J. Phys.
  Soc. Jap.}\ }\textbf {\bibinfo {volume} {74}},\ \bibinfo {pages} {1103}
  (\bibinfo {year} {2005})}\BibitemShut {NoStop}%
\bibitem [{\citenamefont {Balicas}\ \emph {et~al.}(2005)\citenamefont
  {Balicas}, \citenamefont {Nakatsuji}, \citenamefont {Lee}, \citenamefont
  {Schlottmann}, \citenamefont {Murphy},\ and\ \citenamefont
  {Fisk}}]{Balicas2005}%
  \BibitemOpen
  \bibfield  {author} {\bibinfo {author} {\bibfnamefont {L.}~\bibnamefont
  {Balicas}}, \bibinfo {author} {\bibfnamefont {S.}~\bibnamefont {Nakatsuji}},
  \bibinfo {author} {\bibfnamefont {H.}~\bibnamefont {Lee}}, \bibinfo {author}
  {\bibfnamefont {P.}~\bibnamefont {Schlottmann}}, \bibinfo {author}
  {\bibfnamefont {T.~P.}\ \bibnamefont {Murphy}}, \ and\ \bibinfo {author}
  {\bibfnamefont {Z.}~\bibnamefont {Fisk}},\ }\href@noop {} {\bibfield
  {journal} {\bibinfo  {journal} {Phys. Rev. B}\ }\textbf {\bibinfo {volume}
  {72}},\ \bibinfo {pages} {064422} (\bibinfo {year} {2005})}\BibitemShut
  {NoStop}%
\bibitem [{\citenamefont {Zhao}\ \emph {et~al.}(2016)\citenamefont {Zhao},
  \citenamefont {Yelland}, \citenamefont {Bruin}, \citenamefont {Sheikin},
  \citenamefont {Canfield}, \citenamefont {Fritsch}, \citenamefont {Sakai},
  \citenamefont {Mackenzie},\ and\ \citenamefont {Hicks}}]{Zhao2016}%
  \BibitemOpen
  \bibfield  {author} {\bibinfo {author} {\bibfnamefont {L.}~\bibnamefont
  {Zhao}}, \bibinfo {author} {\bibfnamefont {E.~A.}\ \bibnamefont {Yelland}},
  \bibinfo {author} {\bibfnamefont {J.~A.}\ \bibnamefont {Bruin}}, \bibinfo
  {author} {\bibfnamefont {I.}~\bibnamefont {Sheikin}}, \bibinfo {author}
  {\bibfnamefont {P.~C.}\ \bibnamefont {Canfield}}, \bibinfo {author}
  {\bibfnamefont {V.}~\bibnamefont {Fritsch}}, \bibinfo {author} {\bibfnamefont
  {H.}~\bibnamefont {Sakai}}, \bibinfo {author} {\bibfnamefont {A.~P.}\
  \bibnamefont {Mackenzie}}, \ and\ \bibinfo {author} {\bibfnamefont {C.~W.}\
  \bibnamefont {Hicks}},\ }\href@noop {} {\bibfield  {journal} {\bibinfo
  {journal} {Phys. Rev. B}\ }\textbf {\bibinfo {volume} {93}},\ \bibinfo
  {pages} {195124} (\bibinfo {year} {2016})}\BibitemShut {NoStop}%
\bibitem [{\citenamefont {Marcus}\ \emph {et~al.}(2018)\citenamefont {Marcus},
  \citenamefont {Kim}, \citenamefont {Tutmaher}, \citenamefont
  {Rodriguez-Rivera}, \citenamefont {Birk}, \citenamefont {Niedermeyer},
  \citenamefont {Lee}, \citenamefont {Fisk},\ and\ \citenamefont
  {Broholm}}]{Marcus2018}%
  \BibitemOpen
  \bibfield  {author} {\bibinfo {author} {\bibfnamefont {G.~G.}\ \bibnamefont
  {Marcus}}, \bibinfo {author} {\bibfnamefont {D.-J.}\ \bibnamefont {Kim}},
  \bibinfo {author} {\bibfnamefont {J.~A.}\ \bibnamefont {Tutmaher}}, \bibinfo
  {author} {\bibfnamefont {J.~A.}\ \bibnamefont {Rodriguez-Rivera}}, \bibinfo
  {author} {\bibfnamefont {J.~O.}\ \bibnamefont {Birk}}, \bibinfo {author}
  {\bibfnamefont {C.}~\bibnamefont {Niedermeyer}}, \bibinfo {author}
  {\bibfnamefont {H.}~\bibnamefont {Lee}}, \bibinfo {author} {\bibfnamefont
  {Z.}~\bibnamefont {Fisk}}, \ and\ \bibinfo {author} {\bibfnamefont {C.~L.}\
  \bibnamefont {Broholm}},\ }\href@noop {} {\bibfield  {journal} {\bibinfo
  {journal} {Phys. Rev. Lett,}\ }\textbf {\bibinfo {volume} {120}},\ \bibinfo
  {pages} {097201} (\bibinfo {year} {2018})}\BibitemShut {NoStop}%
\bibitem [{\citenamefont {Park}\ \emph
  {et~al.}(2018{\natexlab{a}})\citenamefont {Park}, \citenamefont {Sakai},
  \citenamefont {Erten}, \citenamefont {Mackenzie},\ and\ \citenamefont
  {Hicks}}]{Park2018}%
  \BibitemOpen
  \bibfield  {author} {\bibinfo {author} {\bibfnamefont {J.}~\bibnamefont
  {Park}}, \bibinfo {author} {\bibfnamefont {H.}~\bibnamefont {Sakai}},
  \bibinfo {author} {\bibfnamefont {O.}~\bibnamefont {Erten}}, \bibinfo
  {author} {\bibfnamefont {A.~P.}\ \bibnamefont {Mackenzie}}, \ and\ \bibinfo
  {author} {\bibfnamefont {C.~W.}\ \bibnamefont {Hicks}},\ }\href@noop {}
  {\bibfield  {journal} {\bibinfo  {journal} {Phys. Rev. B}\ }\textbf {\bibinfo
  {volume} {97}},\ \bibinfo {pages} {024411} (\bibinfo {year}
  {2018}{\natexlab{a}})}\BibitemShut {NoStop}%
\bibitem [{\citenamefont {Park}\ \emph
  {et~al.}(2018{\natexlab{b}})\citenamefont {Park}, \citenamefont {Sakai},
  \citenamefont {Mackenzie},\ and\ \citenamefont {Hicks}}]{Park2018a}%
  \BibitemOpen
  \bibfield  {author} {\bibinfo {author} {\bibfnamefont {J.}~\bibnamefont
  {Park}}, \bibinfo {author} {\bibfnamefont {H.}~\bibnamefont {Sakai}},
  \bibinfo {author} {\bibfnamefont {A.~P.}\ \bibnamefont {Mackenzie}}, \ and\
  \bibinfo {author} {\bibfnamefont {C.~W.}\ \bibnamefont {Hicks}},\ }\href@noop
  {} {\bibfield  {journal} {\bibinfo  {journal} {Phys. Rev. B}\ }\textbf
  {\bibinfo {volume} {98}},\ \bibinfo {pages} {024426} (\bibinfo {year}
  {2018}{\natexlab{b}})}\BibitemShut {NoStop}%
\bibitem [{\citenamefont {Thamizhavel}\ \emph {et~al.}(2003)\citenamefont
  {Thamizhavel}, \citenamefont {Takeuchi}, \citenamefont {Okubo}, \citenamefont
  {Yamada}, \citenamefont {Asai}, \citenamefont {Kirita}, \citenamefont
  {Galatanu}, \citenamefont {Yamamoto}, \citenamefont {Ebihara}, \citenamefont
  {Inada}, \citenamefont {Settai},\ and\ \citenamefont
  {\={O}nuki}}]{Thamizhavel2003}%
  \BibitemOpen
  \bibfield  {author} {\bibinfo {author} {\bibfnamefont {A.}~\bibnamefont
  {Thamizhavel}}, \bibinfo {author} {\bibfnamefont {T.}~\bibnamefont
  {Takeuchi}}, \bibinfo {author} {\bibfnamefont {T.}~\bibnamefont {Okubo}},
  \bibinfo {author} {\bibfnamefont {M.}~\bibnamefont {Yamada}}, \bibinfo
  {author} {\bibfnamefont {R.}~\bibnamefont {Asai}}, \bibinfo {author}
  {\bibfnamefont {S.}~\bibnamefont {Kirita}}, \bibinfo {author} {\bibfnamefont
  {A.}~\bibnamefont {Galatanu}}, \bibinfo {author} {\bibfnamefont
  {E.}~\bibnamefont {Yamamoto}}, \bibinfo {author} {\bibfnamefont
  {T.}~\bibnamefont {Ebihara}}, \bibinfo {author} {\bibfnamefont
  {Y.}~\bibnamefont {Inada}}, \bibinfo {author} {\bibfnamefont
  {R.}~\bibnamefont {Settai}}, \ and\ \bibinfo {author} {\bibfnamefont
  {Y.}~\bibnamefont {\={O}nuki}},\ }\href@noop {} {\bibfield  {journal}
  {\bibinfo  {journal} {Phys. Rev. B}\ }\textbf {\bibinfo {volume} {68}},\
  \bibinfo {pages} {054427} (\bibinfo {year} {2003})}\BibitemShut {NoStop}%
\bibitem [{\citenamefont {Chattopadhyay}\ \emph {et~al.}(1994)\citenamefont
  {Chattopadhyay}, \citenamefont {Burlet}, \citenamefont {Rossat-Mignod},
  \citenamefont {Bartholin}, \citenamefont {Vettier},\ and\ \citenamefont
  {Vogt}}]{Chattopadhyay1994}%
  \BibitemOpen
  \bibfield  {author} {\bibinfo {author} {\bibfnamefont {T.}~\bibnamefont
  {Chattopadhyay}}, \bibinfo {author} {\bibfnamefont {P.}~\bibnamefont
  {Burlet}}, \bibinfo {author} {\bibfnamefont {J.}~\bibnamefont
  {Rossat-Mignod}}, \bibinfo {author} {\bibfnamefont {H.}~\bibnamefont
  {Bartholin}}, \bibinfo {author} {\bibfnamefont {C.}~\bibnamefont {Vettier}},
  \ and\ \bibinfo {author} {\bibfnamefont {O.}~\bibnamefont {Vogt}},\ }\href
  {\doibase 10.1103/PhysRevB.49.15096} {\bibfield  {journal} {\bibinfo
  {journal} {Phys. Rev. B}\ }\textbf {\bibinfo {volume} {49}},\ \bibinfo
  {pages} {15096} (\bibinfo {year} {1994})}\BibitemShut {NoStop}%
\bibitem [{\citenamefont {Borzi}\ \emph {et~al.}(2007)\citenamefont {Borzi},
  \citenamefont {Grigera}, \citenamefont {Farrell}, \citenamefont {Perry},
  \citenamefont {Lister}, \citenamefont {Lee}, \citenamefont {Tennant},
  \citenamefont {Maeno},\ and\ \citenamefont {Mackenzie}}]{Borzi2007}%
  \BibitemOpen
  \bibfield  {author} {\bibinfo {author} {\bibfnamefont {R.~A.}\ \bibnamefont
  {Borzi}}, \bibinfo {author} {\bibfnamefont {S.~A.}\ \bibnamefont {Grigera}},
  \bibinfo {author} {\bibfnamefont {J.}~\bibnamefont {Farrell}}, \bibinfo
  {author} {\bibfnamefont {R.~S.}\ \bibnamefont {Perry}}, \bibinfo {author}
  {\bibfnamefont {S.~J.~S.}\ \bibnamefont {Lister}}, \bibinfo {author}
  {\bibfnamefont {S.~L.}\ \bibnamefont {Lee}}, \bibinfo {author} {\bibfnamefont
  {D.~A.}\ \bibnamefont {Tennant}}, \bibinfo {author} {\bibfnamefont
  {Y.}~\bibnamefont {Maeno}}, \ and\ \bibinfo {author} {\bibfnamefont {A.~P.}\
  \bibnamefont {Mackenzie}},\ }\href {\doibase 10.1126/science.1134796}
  {\bibfield  {journal} {\bibinfo  {journal} {Science}\ }\textbf {\bibinfo
  {volume} {315}},\ \bibinfo {pages} {214} (\bibinfo {year}
  {2007})}\BibitemShut {NoStop}%
\bibitem [{\citenamefont {Lester}\ \emph {et~al.}(2015)\citenamefont {Lester},
  \citenamefont {Ramos}, \citenamefont {Perry}, \citenamefont {Croft},
  \citenamefont {Bewley}, \citenamefont {Guidi}, \citenamefont {Manuel},
  \citenamefont {Khalyavin}, \citenamefont {Forgan},\ and\ \citenamefont
  {Hayden}}]{Lester2015}%
  \BibitemOpen
  \bibfield  {author} {\bibinfo {author} {\bibfnamefont {C.}~\bibnamefont
  {Lester}}, \bibinfo {author} {\bibfnamefont {S.}~\bibnamefont {Ramos}},
  \bibinfo {author} {\bibfnamefont {R.}~\bibnamefont {Perry}}, \bibinfo
  {author} {\bibfnamefont {T.}~\bibnamefont {Croft}}, \bibinfo {author}
  {\bibfnamefont {R.}~\bibnamefont {Bewley}}, \bibinfo {author} {\bibfnamefont
  {T.}~\bibnamefont {Guidi}}, \bibinfo {author} {\bibfnamefont
  {P.}~\bibnamefont {Manuel}}, \bibinfo {author} {\bibfnamefont
  {D.}~\bibnamefont {Khalyavin}}, \bibinfo {author} {\bibfnamefont
  {E.}~\bibnamefont {Forgan}}, \ and\ \bibinfo {author} {\bibfnamefont
  {S.}~\bibnamefont {Hayden}},\ }\href@noop {} {\bibfield  {journal} {\bibinfo
  {journal} {Nat. Mater.}\ }\textbf {\bibinfo {volume} {14}},\ \bibinfo {pages}
  {373} (\bibinfo {year} {2015})}\BibitemShut {NoStop}%
\bibitem [{\citenamefont {Mun}\ \emph {et~al.}(2010)\citenamefont {Mun},
  \citenamefont {Budko}, \citenamefont {Kreyssig},\ and\ \citenamefont
  {Canfield}}]{Mun2010}%
  \BibitemOpen
  \bibfield  {author} {\bibinfo {author} {\bibfnamefont {E.}~\bibnamefont
  {Mun}}, \bibinfo {author} {\bibfnamefont {S.~L.}\ \bibnamefont {Budko}},
  \bibinfo {author} {\bibfnamefont {A.}~\bibnamefont {Kreyssig}}, \ and\
  \bibinfo {author} {\bibfnamefont {P.~C.}\ \bibnamefont {Canfield}},\
  }\href@noop {} {\bibfield  {journal} {\bibinfo  {journal} {Phys. Rev. B}\
  }\textbf {\bibinfo {volume} {82}},\ \bibinfo {pages} {054424} (\bibinfo
  {year} {2010})}\BibitemShut {NoStop}%
\bibitem [{\citenamefont {Knafo}\ \emph {et~al.}(2010)\citenamefont {Knafo},
  \citenamefont {Aoki}, \citenamefont {Vignolles}, \citenamefont {Vignolle},
  \citenamefont {Klein}, \citenamefont {Jaudet}, \citenamefont {Villaume},
  \citenamefont {Proust},\ and\ \citenamefont {Flouquet}}]{Knafo2010}%
  \BibitemOpen
  \bibfield  {author} {\bibinfo {author} {\bibfnamefont {W.}~\bibnamefont
  {Knafo}}, \bibinfo {author} {\bibfnamefont {D.}~\bibnamefont {Aoki}},
  \bibinfo {author} {\bibfnamefont {D.}~\bibnamefont {Vignolles}}, \bibinfo
  {author} {\bibfnamefont {B.}~\bibnamefont {Vignolle}}, \bibinfo {author}
  {\bibfnamefont {Y.}~\bibnamefont {Klein}}, \bibinfo {author} {\bibfnamefont
  {C.}~\bibnamefont {Jaudet}}, \bibinfo {author} {\bibfnamefont
  {A.}~\bibnamefont {Villaume}}, \bibinfo {author} {\bibfnamefont
  {C.}~\bibnamefont {Proust}}, \ and\ \bibinfo {author} {\bibfnamefont
  {J.}~\bibnamefont {Flouquet}},\ }\href@noop {} {\bibfield  {journal}
  {\bibinfo  {journal} {Phys. Rev. B}\ }\textbf {\bibinfo {volume} {81}},\
  \bibinfo {pages} {094403} (\bibinfo {year} {2010})}\BibitemShut {NoStop}%
\bibitem [{\citenamefont {Budko}\ \emph {et~al.}(2007)\citenamefont {Budko},
  \citenamefont {Canfield}, \citenamefont {Avila},\ and\ \citenamefont
  {Takabatake}}]{Bud2007}%
  \BibitemOpen
  \bibfield  {author} {\bibinfo {author} {\bibfnamefont {S.~L.}\ \bibnamefont
  {Budko}}, \bibinfo {author} {\bibfnamefont {P.~C.}\ \bibnamefont {Canfield}},
  \bibinfo {author} {\bibfnamefont {M.~A.}\ \bibnamefont {Avila}}, \ and\
  \bibinfo {author} {\bibfnamefont {T.}~\bibnamefont {Takabatake}},\
  }\href@noop {} {\bibfield  {journal} {\bibinfo  {journal} {Phys. Rev. B}\
  }\textbf {\bibinfo {volume} {75}},\ \bibinfo {pages} {094433} (\bibinfo
  {year} {2007})}\BibitemShut {NoStop}%
\bibitem [{\citenamefont {Canfield}\ \emph {et~al.}(2016)\citenamefont
  {Canfield}, \citenamefont {Kong}, \citenamefont {Kaluarachchi},\ and\
  \citenamefont {Jo}}]{Canfield2016}%
  \BibitemOpen
  \bibfield  {author} {\bibinfo {author} {\bibfnamefont {P.~C.}\ \bibnamefont
  {Canfield}}, \bibinfo {author} {\bibfnamefont {T.}~\bibnamefont {Kong}},
  \bibinfo {author} {\bibfnamefont {U.~S.}\ \bibnamefont {Kaluarachchi}}, \
  and\ \bibinfo {author} {\bibfnamefont {N.~H.}\ \bibnamefont {Jo}},\
  }\href@noop {} {\bibfield  {journal} {\bibinfo  {journal} {Phil. Mag.}\
  }\textbf {\bibinfo {volume} {96}},\ \bibinfo {pages} {84} (\bibinfo {year}
  {2016})}\BibitemShut {NoStop}%
\bibitem [{\citenamefont {Canfield}(2020)}]{Canfield2019}%
  \BibitemOpen
  \bibfield  {author} {\bibinfo {author} {\bibfnamefont {P.~C.}\ \bibnamefont
  {Canfield}},\ }\href@noop {} {\bibfield  {journal} {\bibinfo  {journal} {Rep.
  Prog. Phys.}\ }\textbf {\bibinfo {volume} {83}},\ \bibinfo {pages} {016501}
  (\bibinfo {year} {2020})}\BibitemShut {NoStop}%
\bibitem [{\citenamefont {Fisher}(1962)}]{Fisher1962}%
  \BibitemOpen
  \bibfield  {author} {\bibinfo {author} {\bibfnamefont {M.~E.}\ \bibnamefont
  {Fisher}},\ }\href@noop {} {\bibfield  {journal} {\bibinfo  {journal} {Phil.
  Mag.}\ }\textbf {\bibinfo {volume} {7}},\ \bibinfo {pages} {1731} (\bibinfo
  {year} {1962})}\BibitemShut {NoStop}%
\bibitem [{\citenamefont {Ghosh}\ \emph {et~al.}(2020)\citenamefont {Ghosh},
  \citenamefont {Br{\"u}ckner}, \citenamefont {Nikitin}, \citenamefont
  {Grinenko}, \citenamefont {Elender}, \citenamefont {Mackenzie}, \citenamefont
  {Luetkens}, \citenamefont {Klauss},\ and\ \citenamefont {Hicks}}]{Ghosh2020}%
  \BibitemOpen
  \bibfield  {author} {\bibinfo {author} {\bibfnamefont {S.}~\bibnamefont
  {Ghosh}}, \bibinfo {author} {\bibfnamefont {F.}~\bibnamefont {Br{\"u}ckner}},
  \bibinfo {author} {\bibfnamefont {A.}~\bibnamefont {Nikitin}}, \bibinfo
  {author} {\bibfnamefont {V.}~\bibnamefont {Grinenko}}, \bibinfo {author}
  {\bibfnamefont {M.}~\bibnamefont {Elender}}, \bibinfo {author} {\bibfnamefont
  {A.~P.}\ \bibnamefont {Mackenzie}}, \bibinfo {author} {\bibfnamefont
  {H.}~\bibnamefont {Luetkens}}, \bibinfo {author} {\bibfnamefont {H.-H.}\
  \bibnamefont {Klauss}}, \ and\ \bibinfo {author} {\bibfnamefont {C.~W.}\
  \bibnamefont {Hicks}},\ }\href@noop {} {\bibfield  {journal} {\bibinfo
  {journal} {Rev. Sci. Instrum.}\ }\textbf {\bibinfo {volume} {91}},\ \bibinfo
  {pages} {103902} (\bibinfo {year} {2020})}\BibitemShut {NoStop}%
\bibitem [{\citenamefont {Chapon}\ \emph {et~al.}(2011)\citenamefont {Chapon},
  \citenamefont {Manuel}, \citenamefont {Radaelli}, \citenamefont {Benson},
  \citenamefont {Perrott}, \citenamefont {Ansell}, \citenamefont {Rhodes},
  \citenamefont {Raspino}, \citenamefont {Duxbury}, \citenamefont {Spill},\
  and\ \citenamefont {Norris}}]{Chapon2011}%
  \BibitemOpen
  \bibfield  {author} {\bibinfo {author} {\bibfnamefont {L.~C.}\ \bibnamefont
  {Chapon}}, \bibinfo {author} {\bibfnamefont {P.}~\bibnamefont {Manuel}},
  \bibinfo {author} {\bibfnamefont {P.~G.}\ \bibnamefont {Radaelli}}, \bibinfo
  {author} {\bibfnamefont {C.}~\bibnamefont {Benson}}, \bibinfo {author}
  {\bibfnamefont {L.}~\bibnamefont {Perrott}}, \bibinfo {author} {\bibfnamefont
  {S.}~\bibnamefont {Ansell}}, \bibinfo {author} {\bibfnamefont {N.~J.}\
  \bibnamefont {Rhodes}}, \bibinfo {author} {\bibfnamefont {D.}~\bibnamefont
  {Raspino}}, \bibinfo {author} {\bibfnamefont {D.}~\bibnamefont {Duxbury}},
  \bibinfo {author} {\bibfnamefont {E.}~\bibnamefont {Spill}}, \ and\ \bibinfo
  {author} {\bibfnamefont {J.}~\bibnamefont {Norris}},\ }\href@noop {}
  {\bibfield  {journal} {\bibinfo  {journal} {Neutron News}\ }\textbf {\bibinfo
  {volume} {22}},\ \bibinfo {pages} {22} (\bibinfo {year} {2011})}\BibitemShut
  {NoStop}%
\bibitem [{\citenamefont {Arnold}\ \emph {et~al.}(2014)\citenamefont {Arnold},
  \citenamefont {Bilheux}, \citenamefont {Borreguero}, \citenamefont {Buts},
  \citenamefont {Campbell}, \citenamefont {Chapon}, \citenamefont {Doucet},
  \citenamefont {Draper}, \citenamefont {{Ferraz Leal}}, \citenamefont {Gigg},
  \citenamefont {Lynch}, \citenamefont {Markvardsen}, \citenamefont
  {Mikkelson}, \citenamefont {Mikkelson}, \citenamefont {Miller}, \citenamefont
  {Palmen}, \citenamefont {Parker}, \citenamefont {Passos}, \citenamefont
  {Perring}, \citenamefont {Peterson}, \citenamefont {Ren}, \citenamefont
  {Reuter}, \citenamefont {Savici}, \citenamefont {Taylor}, \citenamefont
  {Taylor}, \citenamefont {Tolchenov}, \citenamefont {Zhou},\ and\
  \citenamefont {Zikovsky}}]{Mantid}%
  \BibitemOpen
  \bibfield  {author} {\bibinfo {author} {\bibfnamefont {O.}~\bibnamefont
  {Arnold}}, \bibinfo {author} {\bibfnamefont {J.}~\bibnamefont {Bilheux}},
  \bibinfo {author} {\bibfnamefont {J.}~\bibnamefont {Borreguero}}, \bibinfo
  {author} {\bibfnamefont {A.}~\bibnamefont {Buts}}, \bibinfo {author}
  {\bibfnamefont {S.}~\bibnamefont {Campbell}}, \bibinfo {author}
  {\bibfnamefont {L.}~\bibnamefont {Chapon}}, \bibinfo {author} {\bibfnamefont
  {M.}~\bibnamefont {Doucet}}, \bibinfo {author} {\bibfnamefont
  {N.}~\bibnamefont {Draper}}, \bibinfo {author} {\bibfnamefont
  {R.}~\bibnamefont {{Ferraz Leal}}}, \bibinfo {author} {\bibfnamefont
  {M.}~\bibnamefont {Gigg}}, \bibinfo {author} {\bibfnamefont {V.}~\bibnamefont
  {Lynch}}, \bibinfo {author} {\bibfnamefont {A.}~\bibnamefont {Markvardsen}},
  \bibinfo {author} {\bibfnamefont {D.}~\bibnamefont {Mikkelson}}, \bibinfo
  {author} {\bibfnamefont {R.}~\bibnamefont {Mikkelson}}, \bibinfo {author}
  {\bibfnamefont {R.}~\bibnamefont {Miller}}, \bibinfo {author} {\bibfnamefont
  {K.}~\bibnamefont {Palmen}}, \bibinfo {author} {\bibfnamefont
  {P.}~\bibnamefont {Parker}}, \bibinfo {author} {\bibfnamefont
  {G.}~\bibnamefont {Passos}}, \bibinfo {author} {\bibfnamefont
  {T.}~\bibnamefont {Perring}}, \bibinfo {author} {\bibfnamefont
  {P.}~\bibnamefont {Peterson}}, \bibinfo {author} {\bibfnamefont
  {S.}~\bibnamefont {Ren}}, \bibinfo {author} {\bibfnamefont {M.}~\bibnamefont
  {Reuter}}, \bibinfo {author} {\bibfnamefont {A.}~\bibnamefont {Savici}},
  \bibinfo {author} {\bibfnamefont {J.}~\bibnamefont {Taylor}}, \bibinfo
  {author} {\bibfnamefont {R.}~\bibnamefont {Taylor}}, \bibinfo {author}
  {\bibfnamefont {R.}~\bibnamefont {Tolchenov}}, \bibinfo {author}
  {\bibfnamefont {W.}~\bibnamefont {Zhou}}, \ and\ \bibinfo {author}
  {\bibfnamefont {J.}~\bibnamefont {Zikovsky}},\ }\href
  {https://www.sciencedirect.com/science/article/pii/S0168900214008729}
  {\bibfield  {journal} {\bibinfo  {journal} {Nuc. Inst. Meth. Phys. A}\
  }\textbf {\bibinfo {volume} {764}},\ \bibinfo {pages} {156} (\bibinfo {year}
  {2014})}\BibitemShut {NoStop}%
\bibitem [{\citenamefont {Pet{\v{r}}{\'\i}{\v{c}}ek}\ \emph
  {et~al.}(2014)\citenamefont {Pet{\v{r}}{\'\i}{\v{c}}ek}, \citenamefont
  {Du{\v{s}}ek},\ and\ \citenamefont {Palatinus}}]{Petvrivcek2014}%
  \BibitemOpen
  \bibfield  {author} {\bibinfo {author} {\bibfnamefont {V.}~\bibnamefont
  {Pet{\v{r}}{\'\i}{\v{c}}ek}}, \bibinfo {author} {\bibfnamefont
  {M.}~\bibnamefont {Du{\v{s}}ek}}, \ and\ \bibinfo {author} {\bibfnamefont
  {L.}~\bibnamefont {Palatinus}},\ }\href@noop {} {\bibfield  {journal}
  {\bibinfo  {journal} {Z. Kristallogr. Cryst. Mater.}\ }\textbf {\bibinfo
  {volume} {229}},\ \bibinfo {pages} {345} (\bibinfo {year}
  {2014})}\BibitemShut {NoStop}%
\bibitem [{MVI()}]{MVISUALIZE}%
  \BibitemOpen
  \href@noop {} {}\bibinfo {note} {The mCif files may be viewed using the
  MVISUALIZE software at www.cryst.ehu.es}\BibitemShut {NoStop}%
\bibitem [{\citenamefont {Perez-Mato}\ \emph {et~al.}(2015)\citenamefont
  {Perez-Mato}, \citenamefont {Gallego}, \citenamefont {Tasci}, \citenamefont
  {Elcoro}, \citenamefont {de~la Flor},\ and\ \citenamefont
  {Aroyo}}]{Perez-Mato2015}%
  \BibitemOpen
  \bibfield  {author} {\bibinfo {author} {\bibfnamefont {J.}~\bibnamefont
  {Perez-Mato}}, \bibinfo {author} {\bibfnamefont {S.}~\bibnamefont {Gallego}},
  \bibinfo {author} {\bibfnamefont {E.}~\bibnamefont {Tasci}}, \bibinfo
  {author} {\bibfnamefont {L.}~\bibnamefont {Elcoro}}, \bibinfo {author}
  {\bibfnamefont {G.}~\bibnamefont {de~la Flor}}, \ and\ \bibinfo {author}
  {\bibfnamefont {M.}~\bibnamefont {Aroyo}},\ }\href {\doibase
  10.1146/annurev-matsci-070214-021008} {\bibfield  {journal} {\bibinfo
  {journal} {Annual Review of Materials Research}\ }\textbf {\bibinfo {volume}
  {45}},\ \bibinfo {pages} {217} (\bibinfo {year} {2015})}\BibitemShut
  {NoStop}%
\bibitem [{SI()}]{SI}%
  \BibitemOpen
  \href@noop {} {}\bibinfo {note} {See Supplemental Material at \textit{link to
  be inserted} for Cif and mCif files.}\BibitemShut {Stop}%
\bibitem [{\citenamefont {Sologub}\ \emph {et~al.}(1994)\citenamefont
  {Sologub}, \citenamefont {Hiebl}, \citenamefont {Rogl}, \citenamefont
  {No{\"e}l},\ and\ \citenamefont {Bodak}}]{Sologub1994}%
  \BibitemOpen
  \bibfield  {author} {\bibinfo {author} {\bibfnamefont {O.}~\bibnamefont
  {Sologub}}, \bibinfo {author} {\bibfnamefont {K.}~\bibnamefont {Hiebl}},
  \bibinfo {author} {\bibfnamefont {P.}~\bibnamefont {Rogl}}, \bibinfo {author}
  {\bibfnamefont {H.}~\bibnamefont {No{\"e}l}}, \ and\ \bibinfo {author}
  {\bibfnamefont {O.}~\bibnamefont {Bodak}},\ }\href@noop {} {\bibfield
  {journal} {\bibinfo  {journal} {J. Alloys Compd.}\ }\textbf {\bibinfo
  {volume} {210}},\ \bibinfo {pages} {153} (\bibinfo {year}
  {1994})}\BibitemShut {NoStop}%
\bibitem [{\citenamefont {Becker}\ and\ \citenamefont
  {Coppens}(1974)}]{Becker1974}%
  \BibitemOpen
  \bibfield  {author} {\bibinfo {author} {\bibfnamefont {P.~J.}\ \bibnamefont
  {Becker}}\ and\ \bibinfo {author} {\bibfnamefont {P.}~\bibnamefont
  {Coppens}},\ }\href@noop {} {\bibfield  {journal} {\bibinfo  {journal} {Acta
  Cryst. A}\ }\textbf {\bibinfo {volume} {30}},\ \bibinfo {pages} {129}
  (\bibinfo {year} {1974})}\BibitemShut {NoStop}%
\bibitem [{\citenamefont {Hatch}\ and\ \citenamefont
  {Stokes}(2003)}]{Hatch2003}%
  \BibitemOpen
  \bibfield  {author} {\bibinfo {author} {\bibfnamefont {D.~M.}\ \bibnamefont
  {Hatch}}\ and\ \bibinfo {author} {\bibfnamefont {H.~T.}\ \bibnamefont
  {Stokes}},\ }\href {\doibase 10.1107/S0021889803005946} {\bibfield  {journal}
  {\bibinfo  {journal} {J. App. Cryst.}\ }\textbf {\bibinfo {volume} {36}},\
  \bibinfo {pages} {951} (\bibinfo {year} {2003})}\BibitemShut {NoStop}%
\bibitem [{\citenamefont {Stokes}\ \emph {et~al.}()\citenamefont {Stokes},
  \citenamefont {Hatch},\ and\ \citenamefont {Campbell}}]{stokes2016}%
  \BibitemOpen
  \bibfield  {author} {\bibinfo {author} {\bibfnamefont {H.~T.}\ \bibnamefont
  {Stokes}}, \bibinfo {author} {\bibfnamefont {D.~M.}\ \bibnamefont {Hatch}}, \
  and\ \bibinfo {author} {\bibfnamefont {B.~J.}\ \bibnamefont {Campbell}},\
  }\href@noop {} {}\bibinfo {note} {ISOTROPY Software Suite,
  iso.byu.edu}\BibitemShut {NoStop}%
\bibitem [{\citenamefont {Anand}\ \emph {et~al.}(2018)\citenamefont {Anand},
  \citenamefont {Hillier}, \citenamefont {Adroja}, \citenamefont {Khalyavin},
  \citenamefont {Manuel}, \citenamefont {Andre}, \citenamefont {Rols},\ and\
  \citenamefont {Koza}}]{Anand2018}%
  \BibitemOpen
  \bibfield  {author} {\bibinfo {author} {\bibfnamefont {V.~K.}\ \bibnamefont
  {Anand}}, \bibinfo {author} {\bibfnamefont {A.~D.}\ \bibnamefont {Hillier}},
  \bibinfo {author} {\bibfnamefont {D.~T.}\ \bibnamefont {Adroja}}, \bibinfo
  {author} {\bibfnamefont {D.~D.}\ \bibnamefont {Khalyavin}}, \bibinfo {author}
  {\bibfnamefont {P.}~\bibnamefont {Manuel}}, \bibinfo {author} {\bibfnamefont
  {G.}~\bibnamefont {Andre}}, \bibinfo {author} {\bibfnamefont
  {S.}~\bibnamefont {Rols}}, \ and\ \bibinfo {author} {\bibfnamefont {M.~M.}\
  \bibnamefont {Koza}},\ }\href {\doibase 10.1103/PhysRevB.97.184422}
  {\bibfield  {journal} {\bibinfo  {journal} {Phys. Rev. B}\ }\textbf {\bibinfo
  {volume} {97}},\ \bibinfo {pages} {184422} (\bibinfo {year}
  {2018})}\BibitemShut {NoStop}%
\bibitem [{\citenamefont {Hillier}\ \emph {et~al.}(2012)\citenamefont
  {Hillier}, \citenamefont {Adroja}, \citenamefont {Manuel}, \citenamefont
  {Anand}, \citenamefont {Taylor}, \citenamefont {McEwen}, \citenamefont
  {Rainford},\ and\ \citenamefont {Koza}}]{Hillier2012}%
  \BibitemOpen
  \bibfield  {author} {\bibinfo {author} {\bibfnamefont {A.~D.}\ \bibnamefont
  {Hillier}}, \bibinfo {author} {\bibfnamefont {D.~T.}\ \bibnamefont {Adroja}},
  \bibinfo {author} {\bibfnamefont {P.}~\bibnamefont {Manuel}}, \bibinfo
  {author} {\bibfnamefont {V.~K.}\ \bibnamefont {Anand}}, \bibinfo {author}
  {\bibfnamefont {J.~W.}\ \bibnamefont {Taylor}}, \bibinfo {author}
  {\bibfnamefont {K.~A.}\ \bibnamefont {McEwen}}, \bibinfo {author}
  {\bibfnamefont {B.~D.}\ \bibnamefont {Rainford}}, \ and\ \bibinfo {author}
  {\bibfnamefont {M.~M.}\ \bibnamefont {Koza}},\ }\href {\doibase
  10.1103/PhysRevB.85.134405} {\bibfield  {journal} {\bibinfo  {journal} {Phys.
  Rev. B}\ }\textbf {\bibinfo {volume} {85}},\ \bibinfo {pages} {134405}
  (\bibinfo {year} {2012})}\BibitemShut {NoStop}%
\bibitem [{\citenamefont {Mostovoy}(2006)}]{Mostovoy2006}%
  \BibitemOpen
  \bibfield  {author} {\bibinfo {author} {\bibfnamefont {M.}~\bibnamefont
  {Mostovoy}},\ }\href {\doibase 10.1103/PhysRevLett.96.067601} {\bibfield
  {journal} {\bibinfo  {journal} {Phys. Rev. Lett.}\ }\textbf {\bibinfo
  {volume} {96}},\ \bibinfo {pages} {067601} (\bibinfo {year}
  {2006})}\BibitemShut {NoStop}%
\bibitem [{\citenamefont {Princep}\ \emph {et~al.}(2020)\citenamefont
  {Princep}, \citenamefont {Feng}, \citenamefont {Guo}, \citenamefont {Lang},
  \citenamefont {Weng}, \citenamefont {Manuel}, \citenamefont {Khalyavin},
  \citenamefont {Senyshyn}, \citenamefont {Rahn}, \citenamefont {Yuan},
  \citenamefont {Matsushita}, \citenamefont {Blundell}, \citenamefont
  {Yamaura},\ and\ \citenamefont {Boothroyd}}]{Princep2020}%
  \BibitemOpen
  \bibfield  {author} {\bibinfo {author} {\bibfnamefont {A.~J.}\ \bibnamefont
  {Princep}}, \bibinfo {author} {\bibfnamefont {H.~L.}\ \bibnamefont {Feng}},
  \bibinfo {author} {\bibfnamefont {Y.~F.}\ \bibnamefont {Guo}}, \bibinfo
  {author} {\bibfnamefont {F.}~\bibnamefont {Lang}}, \bibinfo {author}
  {\bibfnamefont {H.~M.}\ \bibnamefont {Weng}}, \bibinfo {author}
  {\bibfnamefont {P.}~\bibnamefont {Manuel}}, \bibinfo {author} {\bibfnamefont
  {D.}~\bibnamefont {Khalyavin}}, \bibinfo {author} {\bibfnamefont
  {A.}~\bibnamefont {Senyshyn}}, \bibinfo {author} {\bibfnamefont {M.~C.}\
  \bibnamefont {Rahn}}, \bibinfo {author} {\bibfnamefont {Y.~H.}\ \bibnamefont
  {Yuan}}, \bibinfo {author} {\bibfnamefont {Y.}~\bibnamefont {Matsushita}},
  \bibinfo {author} {\bibfnamefont {S.~J.}\ \bibnamefont {Blundell}}, \bibinfo
  {author} {\bibfnamefont {K.}~\bibnamefont {Yamaura}}, \ and\ \bibinfo
  {author} {\bibfnamefont {A.~T.}\ \bibnamefont {Boothroyd}},\ }\href
  {https://link.aps.org/doi/10.1103/PhysRevB.102.104410} {\bibfield  {journal}
  {\bibinfo  {journal} {Phys. Rev. B}\ }\textbf {\bibinfo {volume} {102}},\
  \bibinfo {pages} {104410} (\bibinfo {year} {2020})}\BibitemShut {NoStop}%
\bibitem [{\citenamefont {Feng}\ \emph {et~al.}(2021)\citenamefont {Feng},
  \citenamefont {Kang}, \citenamefont {Manuel}, \citenamefont {Orlandi},
  \citenamefont {Su}, \citenamefont {Chen}, \citenamefont {Tsujimoto},
  \citenamefont {Hadermann}, \citenamefont {Kotliar}, \citenamefont {Yamaura},
  \citenamefont {McCabe},\ and\ \citenamefont {Greenblatt}}]{Feng2021}%
  \BibitemOpen
  \bibfield  {author} {\bibinfo {author} {\bibfnamefont {H.~L.}\ \bibnamefont
  {Feng}}, \bibinfo {author} {\bibfnamefont {C.-J.}\ \bibnamefont {Kang}},
  \bibinfo {author} {\bibfnamefont {P.}~\bibnamefont {Manuel}}, \bibinfo
  {author} {\bibfnamefont {F.}~\bibnamefont {Orlandi}}, \bibinfo {author}
  {\bibfnamefont {Y.}~\bibnamefont {Su}}, \bibinfo {author} {\bibfnamefont
  {J.}~\bibnamefont {Chen}}, \bibinfo {author} {\bibfnamefont {Y.}~\bibnamefont
  {Tsujimoto}}, \bibinfo {author} {\bibfnamefont {J.}~\bibnamefont
  {Hadermann}}, \bibinfo {author} {\bibfnamefont {G.}~\bibnamefont {Kotliar}},
  \bibinfo {author} {\bibfnamefont {K.}~\bibnamefont {Yamaura}}, \bibinfo
  {author} {\bibfnamefont {E.~E.}\ \bibnamefont {McCabe}}, \ and\ \bibinfo
  {author} {\bibfnamefont {M.}~\bibnamefont {Greenblatt}},\ }\href
  {https://doi.org/10.1021/acs.chemmater.1c01032} {\bibfield  {journal}
  {\bibinfo  {journal} {Chem. Mater.}\ }\textbf {\bibinfo {volume} {33}},\
  \bibinfo {pages} {4188} (\bibinfo {year} {2021})}\BibitemShut {NoStop}%
\bibitem [{\citenamefont {Fawcett}(1988)}]{Fawcett1988}%
  \BibitemOpen
  \bibfield  {author} {\bibinfo {author} {\bibfnamefont {E.}~\bibnamefont
  {Fawcett}},\ }\href {\doibase 10.1103/RevModPhys.60.209} {\bibfield
  {journal} {\bibinfo  {journal} {Rev, Mod. Phys.}\ }\textbf {\bibinfo {volume}
  {60}},\ \bibinfo {pages} {209} (\bibinfo {year} {1988})}\BibitemShut
  {NoStop}%
\bibitem [{\citenamefont {Pratt}\ \emph {et~al.}(2011)\citenamefont {Pratt},
  \citenamefont {Kim}, \citenamefont {Kreyssig}, \citenamefont {Lee},
  \citenamefont {Tucker}, \citenamefont {Thaler}, \citenamefont {Tian},
  \citenamefont {Zarestky}, \citenamefont {Budko}, \citenamefont {Canfield},
  \citenamefont {Harmon}, \citenamefont {Goldman},\ and\ \citenamefont
  {J}}]{Pratt2011}%
  \BibitemOpen
  \bibfield  {author} {\bibinfo {author} {\bibfnamefont {D.~K.}\ \bibnamefont
  {Pratt}}, \bibinfo {author} {\bibfnamefont {M.~G.}\ \bibnamefont {Kim}},
  \bibinfo {author} {\bibfnamefont {A.}~\bibnamefont {Kreyssig}}, \bibinfo
  {author} {\bibfnamefont {Y.~B.}\ \bibnamefont {Lee}}, \bibinfo {author}
  {\bibfnamefont {G.~S.}\ \bibnamefont {Tucker}}, \bibinfo {author}
  {\bibfnamefont {A.}~\bibnamefont {Thaler}}, \bibinfo {author} {\bibfnamefont
  {W.}~\bibnamefont {Tian}}, \bibinfo {author} {\bibfnamefont {J.~L.}\
  \bibnamefont {Zarestky}}, \bibinfo {author} {\bibfnamefont {S.~L.}\
  \bibnamefont {Budko}}, \bibinfo {author} {\bibfnamefont {P.~C.}\ \bibnamefont
  {Canfield}}, \bibinfo {author} {\bibfnamefont {B.~N.}\ \bibnamefont
  {Harmon}}, \bibinfo {author} {\bibfnamefont {A.~I.}\ \bibnamefont {Goldman}},
  \ and\ \bibinfo {author} {\bibfnamefont {M.~R.}\ \bibnamefont {J}},\
  }\href@noop {} {\bibfield  {journal} {\bibinfo  {journal} {Phys. Rev. Lett,}\
  }\textbf {\bibinfo {volume} {106}},\ \bibinfo {pages} {257001} (\bibinfo
  {year} {2011})}\BibitemShut {NoStop}%
\bibitem [{\citenamefont {Lee}\ \emph {et~al.}(1993)\citenamefont {Lee},
  \citenamefont {Rice},\ and\ \citenamefont {Anderson}}]{Lee1993}%
  \BibitemOpen
  \bibfield  {author} {\bibinfo {author} {\bibfnamefont {P.}~\bibnamefont
  {Lee}}, \bibinfo {author} {\bibfnamefont {T.}~\bibnamefont {Rice}}, \ and\
  \bibinfo {author} {\bibfnamefont {P.}~\bibnamefont {Anderson}},\ }\href@noop
  {} {\bibfield  {journal} {\bibinfo  {journal} {Solid State Commun.}\ }\textbf
  {\bibinfo {volume} {88}},\ \bibinfo {pages} {1001} (\bibinfo {year}
  {1993})}\BibitemShut {NoStop}%
\bibitem [{\citenamefont {Fleming}\ \emph {et~al.}(1984)\citenamefont
  {Fleming}, \citenamefont {Moncton}, \citenamefont {Axe},\ and\ \citenamefont
  {Brown}}]{Fleming1984}%
  \BibitemOpen
  \bibfield  {author} {\bibinfo {author} {\bibfnamefont {R.}~\bibnamefont
  {Fleming}}, \bibinfo {author} {\bibfnamefont {D.}~\bibnamefont {Moncton}},
  \bibinfo {author} {\bibfnamefont {J.}~\bibnamefont {Axe}}, \ and\ \bibinfo
  {author} {\bibfnamefont {G.}~\bibnamefont {Brown}},\ }\href@noop {}
  {\bibfield  {journal} {\bibinfo  {journal} {Phys. Rev. B}\ }\textbf {\bibinfo
  {volume} {30}},\ \bibinfo {pages} {1877} (\bibinfo {year}
  {1984})}\BibitemShut {NoStop}%
\bibitem [{\citenamefont {Avci}\ \emph {et~al.}(2014)\citenamefont {Avci},
  \citenamefont {Chmaissem}, \citenamefont {Allred}, \citenamefont
  {Rosenkranz}, \citenamefont {Eremin}, \citenamefont {Chubukov}, \citenamefont
  {Bugaris}, \citenamefont {Chung}, \citenamefont {Kanatzidis}, \citenamefont
  {Castellan}, \citenamefont {Schlueter}, \citenamefont {Claus}, \citenamefont
  {Khalyavin}, \citenamefont {Manuel}, \citenamefont {Daoud-Aladine},\ and\
  \citenamefont {Osborn}}]{Avci2014}%
  \BibitemOpen
  \bibfield  {author} {\bibinfo {author} {\bibfnamefont {S.}~\bibnamefont
  {Avci}}, \bibinfo {author} {\bibfnamefont {O.}~\bibnamefont {Chmaissem}},
  \bibinfo {author} {\bibfnamefont {J.}~\bibnamefont {Allred}}, \bibinfo
  {author} {\bibfnamefont {S.}~\bibnamefont {Rosenkranz}}, \bibinfo {author}
  {\bibfnamefont {I.}~\bibnamefont {Eremin}}, \bibinfo {author} {\bibfnamefont
  {A.~V.}\ \bibnamefont {Chubukov}}, \bibinfo {author} {\bibfnamefont
  {D.}~\bibnamefont {Bugaris}}, \bibinfo {author} {\bibfnamefont
  {D.}~\bibnamefont {Chung}}, \bibinfo {author} {\bibfnamefont {M.~G.}\
  \bibnamefont {Kanatzidis}}, \bibinfo {author} {\bibfnamefont {J.-P.}\
  \bibnamefont {Castellan}}, \bibinfo {author} {\bibfnamefont {J.~A.}\
  \bibnamefont {Schlueter}}, \bibinfo {author} {\bibfnamefont {H.}~\bibnamefont
  {Claus}}, \bibinfo {author} {\bibfnamefont {D.~D.}\ \bibnamefont
  {Khalyavin}}, \bibinfo {author} {\bibfnamefont {P.}~\bibnamefont {Manuel}},
  \bibinfo {author} {\bibfnamefont {A.}~\bibnamefont {Daoud-Aladine}}, \ and\
  \bibinfo {author} {\bibfnamefont {R.}~\bibnamefont {Osborn}},\ }\href@noop {}
  {\bibfield  {journal} {\bibinfo  {journal} {Nat. Commun.}\ }\textbf {\bibinfo
  {volume} {5}},\ \bibinfo {pages} {1} (\bibinfo {year} {2014})}\BibitemShut
  {NoStop}%
\bibitem [{\citenamefont {Allred}\ \emph {et~al.}(2016)\citenamefont {Allred},
  \citenamefont {Taddei}, \citenamefont {Bugaris}, \citenamefont {Krogstad},
  \citenamefont {Lapidus}, \citenamefont {Chung}, \citenamefont {Claus},
  \citenamefont {Kanatzidis}, \citenamefont {Brown}, \citenamefont {Kang},
  \citenamefont {Fernandes}, \citenamefont {Eremin}, \citenamefont
  {Rosenkranz}, \citenamefont {Chmaissem},\ and\ \citenamefont
  {Osborn}}]{Allred2016_ATBK}%
  \BibitemOpen
  \bibfield  {author} {\bibinfo {author} {\bibfnamefont {J.~M.}\ \bibnamefont
  {Allred}}, \bibinfo {author} {\bibfnamefont {K.~M.}\ \bibnamefont {Taddei}},
  \bibinfo {author} {\bibfnamefont {D.~E.}\ \bibnamefont {Bugaris}}, \bibinfo
  {author} {\bibfnamefont {M.~J.}\ \bibnamefont {Krogstad}}, \bibinfo {author}
  {\bibfnamefont {S.~H.}\ \bibnamefont {Lapidus}}, \bibinfo {author}
  {\bibfnamefont {D.~Y.}\ \bibnamefont {Chung}}, \bibinfo {author}
  {\bibfnamefont {H.}~\bibnamefont {Claus}}, \bibinfo {author} {\bibfnamefont
  {M.~G.}\ \bibnamefont {Kanatzidis}}, \bibinfo {author} {\bibfnamefont
  {D.~E.}\ \bibnamefont {Brown}}, \bibinfo {author} {\bibfnamefont
  {J.}~\bibnamefont {Kang}}, \bibinfo {author} {\bibfnamefont {R.~M.}\
  \bibnamefont {Fernandes}}, \bibinfo {author} {\bibfnamefont {I.}~\bibnamefont
  {Eremin}}, \bibinfo {author} {\bibfnamefont {S.}~\bibnamefont {Rosenkranz}},
  \bibinfo {author} {\bibfnamefont {O.}~\bibnamefont {Chmaissem}}, \ and\
  \bibinfo {author} {\bibfnamefont {R.}~\bibnamefont {Osborn}},\ }\href
  {https://doi.org/10.1038/nphys3629} {\bibfield  {journal} {\bibinfo
  {journal} {Nat. Phys.}\ }\textbf {\bibinfo {volume} {12}},\ \bibinfo {pages}
  {493} (\bibinfo {year} {2016})}\BibitemShut {NoStop}%
\bibitem [{\citenamefont {Yumnam}\ \emph {et~al.}(2019)\citenamefont {Yumnam},
  \citenamefont {Chen}, \citenamefont {Zhao}, \citenamefont {Thamizhavel},
  \citenamefont {Dhar},\ and\ \citenamefont {Singh}}]{Yumnam2019}%
  \BibitemOpen
  \bibfield  {author} {\bibinfo {author} {\bibfnamefont {G.}~\bibnamefont
  {Yumnam}}, \bibinfo {author} {\bibfnamefont {Y.}~\bibnamefont {Chen}},
  \bibinfo {author} {\bibfnamefont {Y.}~\bibnamefont {Zhao}}, \bibinfo {author}
  {\bibfnamefont {A.}~\bibnamefont {Thamizhavel}}, \bibinfo {author}
  {\bibfnamefont {S.~K.}\ \bibnamefont {Dhar}}, \ and\ \bibinfo {author}
  {\bibfnamefont {D.~K.}\ \bibnamefont {Singh}},\ }\href@noop {} {\bibfield
  {journal} {\bibinfo  {journal} {Phys. Status Solidi B}\ }\textbf {\bibinfo
  {volume} {13}},\ \bibinfo {pages} {1900304} (\bibinfo {year}
  {2019})}\BibitemShut {NoStop}%
\bibitem [{\citenamefont {Efremov}\ \emph {et~al.}(2019)\citenamefont
  {Efremov}, \citenamefont {Shtyk}, \citenamefont {Rost}, \citenamefont
  {Chamon}, \citenamefont {Mackenzie},\ and\ \citenamefont
  {Betouras}}]{Efremov2018}%
  \BibitemOpen
  \bibfield  {author} {\bibinfo {author} {\bibfnamefont {D.~V.}\ \bibnamefont
  {Efremov}}, \bibinfo {author} {\bibfnamefont {A.}~\bibnamefont {Shtyk}},
  \bibinfo {author} {\bibfnamefont {A.~W.}\ \bibnamefont {Rost}}, \bibinfo
  {author} {\bibfnamefont {C.}~\bibnamefont {Chamon}}, \bibinfo {author}
  {\bibfnamefont {A.~P.}\ \bibnamefont {Mackenzie}}, \ and\ \bibinfo {author}
  {\bibfnamefont {J.~J.}\ \bibnamefont {Betouras}},\ }\href@noop {} {\bibfield
  {journal} {\bibinfo  {journal} {Phys. Rev. Lett,}\ }\textbf {\bibinfo
  {volume} {123}},\ \bibinfo {pages} {207202} (\bibinfo {year}
  {2019})}\BibitemShut {NoStop}%
\bibitem [{\citenamefont {Muir}\ and\ \citenamefont
  {Str{\"o}m-Olsen}(1971)}]{Muir1971}%
  \BibitemOpen
  \bibfield  {author} {\bibinfo {author} {\bibfnamefont {W.}~\bibnamefont
  {Muir}}\ and\ \bibinfo {author} {\bibfnamefont {J.}~\bibnamefont
  {Str{\"o}m-Olsen}},\ }\href@noop {} {\bibfield  {journal} {\bibinfo
  {journal} {Phys. Rev. B}\ }\textbf {\bibinfo {volume} {4}},\ \bibinfo {pages}
  {988} (\bibinfo {year} {1971})}\BibitemShut {NoStop}%
\bibitem [{\citenamefont {Millis}\ and\ \citenamefont
  {Norman}(2007)}]{Millis2007}%
  \BibitemOpen
  \bibfield  {author} {\bibinfo {author} {\bibfnamefont {A.~J.}\ \bibnamefont
  {Millis}}\ and\ \bibinfo {author} {\bibfnamefont {M.~R.}\ \bibnamefont
  {Norman}},\ }\href {\doibase 10.1103/PhysRevB.76.220503} {\bibfield
  {journal} {\bibinfo  {journal} {Phys. Rev. B}\ }\textbf {\bibinfo {volume}
  {76}},\ \bibinfo {pages} {220503} (\bibinfo {year} {2007})}\BibitemShut
  {NoStop}%
\bibitem [{\citenamefont {Knafo}\ \emph {et~al.}(2016)\citenamefont {Knafo},
  \citenamefont {Duc}, \citenamefont {Bourdarot}, \citenamefont {Kuwahara},
  \citenamefont {Nojiri}, \citenamefont {Aoki}, \citenamefont {Billette},
  \citenamefont {Frings}, \citenamefont {Tonon}, \citenamefont
  {Lelievre-Berna}, \citenamefont {Flouquet},\ and\ \citenamefont
  {Regnault}}]{Knafo2016}%
  \BibitemOpen
  \bibfield  {author} {\bibinfo {author} {\bibfnamefont {W.}~\bibnamefont
  {Knafo}}, \bibinfo {author} {\bibfnamefont {F.}~\bibnamefont {Duc}}, \bibinfo
  {author} {\bibfnamefont {F.}~\bibnamefont {Bourdarot}}, \bibinfo {author}
  {\bibfnamefont {K.}~\bibnamefont {Kuwahara}}, \bibinfo {author}
  {\bibfnamefont {H.}~\bibnamefont {Nojiri}}, \bibinfo {author} {\bibfnamefont
  {D.}~\bibnamefont {Aoki}}, \bibinfo {author} {\bibfnamefont {J.}~\bibnamefont
  {Billette}}, \bibinfo {author} {\bibfnamefont {P.}~\bibnamefont {Frings}},
  \bibinfo {author} {\bibfnamefont {X.}~\bibnamefont {Tonon}}, \bibinfo
  {author} {\bibfnamefont {E.}~\bibnamefont {Lelievre-Berna}}, \bibinfo
  {author} {\bibfnamefont {J.}~\bibnamefont {Flouquet}}, \ and\ \bibinfo
  {author} {\bibfnamefont {L.-P.}\ \bibnamefont {Regnault}},\ }\href@noop {}
  {\bibfield  {journal} {\bibinfo  {journal} {Nat. Commun.}\ }\textbf {\bibinfo
  {volume} {7}},\ \bibinfo {pages} {13075} (\bibinfo {year}
  {2016})}\BibitemShut {NoStop}%
\bibitem [{\citenamefont {Aroyo}\ \emph {et~al.}(2006)\citenamefont {Aroyo},
  \citenamefont {Kirov}, \citenamefont {Capillas}, \citenamefont {Perez-Mato},\
  and\ \citenamefont {Wondratschek}}]{Aroyo2006}%
  \BibitemOpen
  \bibfield  {author} {\bibinfo {author} {\bibfnamefont {M.~I.}\ \bibnamefont
  {Aroyo}}, \bibinfo {author} {\bibfnamefont {A.}~\bibnamefont {Kirov}},
  \bibinfo {author} {\bibfnamefont {C.}~\bibnamefont {Capillas}}, \bibinfo
  {author} {\bibfnamefont {J.~M.}\ \bibnamefont {Perez-Mato}}, \ and\ \bibinfo
  {author} {\bibfnamefont {H.}~\bibnamefont {Wondratschek}},\ }\href {\doibase
  10.1107/S0108767305040286} {\bibfield  {journal} {\bibinfo  {journal} {Acta
  Cryst. A}\ }\textbf {\bibinfo {volume} {62}},\ \bibinfo {pages} {115}
  (\bibinfo {year} {2006})}\BibitemShut {NoStop}%
\bibitem [{\citenamefont {Von~Dreele}\ \emph {et~al.}(1982)\citenamefont
  {Von~Dreele}, \citenamefont {Jorgensen},\ and\ \citenamefont
  {Windsor}}]{VonDreele1982}%
  \BibitemOpen
  \bibfield  {author} {\bibinfo {author} {\bibfnamefont {R.~B.}\ \bibnamefont
  {Von~Dreele}}, \bibinfo {author} {\bibfnamefont {J.~D.}\ \bibnamefont
  {Jorgensen}}, \ and\ \bibinfo {author} {\bibfnamefont {C.~G.}\ \bibnamefont
  {Windsor}},\ }\href@noop {} {\bibfield  {journal} {\bibinfo  {journal} {J.
  Appl. Cryst.}\ }\textbf {\bibinfo {volume} {15}},\ \bibinfo {pages} {581}
  (\bibinfo {year} {1982})}\BibitemShut {NoStop}%
\end{thebibliography}
%

\end{document}